\documentclass[aps,pra,reprint,nofootinbib,nobibnotes]{revtex4-2}
\usepackage{overpic}
\usepackage{graphicx}
\usepackage{tikz}
\usetikzlibrary{arrows.meta}
\usepackage{amsmath,amssymb,bm,mathtools}
\usepackage{booktabs}
\usepackage[colorlinks=true,citecolor=blue,linkcolor=red,urlcolor=blue]{hyperref}

\newcommand{\Tr}{\operatorname{Tr}}
\newcommand{\ee}{\mathrm{e}}
\newcommand{\ii}{\mathrm{i}}
\newcommand{\kb}{k_{\mathrm B}}
\newcommand{\id}{\mathbb I}
\newcommand{\epsc}{\hbar\omega_c}
\newcommand{\one}{\bm 1}

\begin{document}

\title{Measurement-Feedback Quantum Information Engine: Coherence–Transition Interference and Correlated Work Statistics}

\author{Yingying Hong$^{1}$}
\author{Dehua Liu$^{1}$}
\author{Jinfeng Wei$^{2}$}
\author{Leilei Yan$^{2}$}\email{llyan@zzu.edu.cn}
\author{Jianhui Wang$^{1}$}\email{wangjianhui@ncu.edu.cn}
\affiliation{$^1$ Department of Physics, Nanchang University,
Nanchang 330031, China\\
$^2$ Key Laboratory of Materials Physics, Ministry of Education,
School of Physics and Microelectronics, Zhengzhou University,
Zhengzhou 450001, China}

\date{\today}

\begin{abstract}
 Measurement and feedback jointly prepare coherence and select finite-time dynamics in quantum information engines. Complementing a companion experimental realization, we develop a mechanism-resolved theory of the resulting work statistics and temporal correlations. The conditional work separates into population transfer and a phase-sensitive coherence--transition interference term. Symmetric full counting statistics maps this interference to an equal-and-opposite half-quantum pair in the work quasiprobability, entering odd moments while leaving even moments fixed by endpoint mixing. An outcome-resolved tilted kernel then propagates these statistics through the correlated measurement record, yielding finite-cycle and fixed-time fluctuation corrections. The same memory reduces the reversible record-reset cost from the one-symbol entropy to the entropy rate. Our results link coherent work statistics, feedback memory, and information thermodynamics.
\end{abstract}

\maketitle

\section{Introduction}
\label{sec:introduction}

Measurement and feedback \cite{Wiseman2010} convert information acquired from a working medium
into controlled energy exchange \cite{Parrondo2015,Koski2014}.  At the quantum scale the measurement is
also a preparation step \cite{Goold2016,Blok2014}: its backaction changes the energy populations \cite{Elouard2017npj},
creates coherence \cite{Manikandan2022,LinSu2021}, and selects the feedback operation
\cite{Sagawa2010,Elouard2018prl}.  If that
operation is performed in finite time \cite{Su2023,Fadler2023}, the prepared coherence and the
nonadiabatic transition amplitudes act simultaneously \cite{Feldmann2003,Wang2019}.  The resulting work
cannot in general be separated into a positive ``friction cost'' and a
monotonic coherence advantage \cite{Rezek2006,Francica2019}.  In particular, the same transition channel
that costs work for a passive Gibbs state may release energy from an active
postmeasurement state \cite{Plastina2014}, while its coherent contribution retains a dynamical
relative phase \cite{Xu2018}.

This local mechanism is hidden by a two-projective-energy-measurement
construction because the first projection removes the initial energy
coherence \cite{Perarnau2017,Campisi2011}.  It can instead be resolved with
symmetric $q=1/2$ full counting statistics (FCS) \cite{Esposito2009,Xu2018}, whose Fourier transform is
generally a signed quasiprobability
\cite{Solinas2015,Perarnau2017,Lostaglio2018,Gherardini2024}.  For the feedback branch studied below, the integer
sector records population transfer and the half-quantum sector records
interference between energy amplitudes that reach the same final state.  This
separation gives a direct test of which work moments contain the measurement
phase \cite{Pei2023,Gherardini2024}.

Repeated feedback cycles add temporal structure \cite{Gong2016,Murashita2017}.  The outcome in one cycle
selects both the state and the branch entering the next, so a stationary
one-cycle distribution does not imply independent work increments \cite{Flindt2010,FeiLiu2022}.  {The outcome-resolved tilted kernel retains
work--next-outcome correlations and their propagation through the outcome
process, and is therefore required for finite-cycle and fixed-time
fluctuations. Correlations in the outcome record also affect the information
ledger} \cite{HorowitzEsposito2014,Strasberg2013}: for a buffered record, reversible reset \cite{Landauer1961,Bennett1982} is governed by the entropy
rate \cite{Shannon1948,CoverThomas} rather than the one-symbol entropy.

{
Solinas and Gasparinetti \cite{Solinas2015}
developed a detector-based formulation of
work statistics for arbitrary initial states,
including the effects of initial coherence.
Xu et al.~\cite{Xu2018} further investigated
phase-dependent coherence contributions to
work quasiprobabilities and fluctuations
using FCS. These studies established
coherence-sensitive work statistics for
driven quantum processes.
Bresque et al.~\cite{Bresque2025} derived
exact finite-time work statistics and
first-passage-time statistics for a lazy
quantum measurement engine through a
run-and-tumble mapping, revealing correlations
across repeated cycles.
Our earlier feedback-driven Otto-like engine
\cite{Liu25} examined finite-time thermodynamic
performance and single-cycle FCS work
fluctuations.
Building on these studies, the present work
jointly propagates coherence-sensitive work
statistics and feedback outcomes through an
outcome-resolved tilted kernel.
Crucially, this kernel retains
work--next-outcome correlations that cannot
generally be reconstructed from the
single-cycle work distribution and the
outcome transition matrix alone.
}

The local coherence–transition interference can be propagated directly through the correlated measurement record.  First, the conditional mean work is separated into population
transfer and a signed coherence--transition interference term.  Symmetric FCS
identifies the latter with an equal-and-opposite half-quantum pair and yields
an odd--even moment separation: the interference enters the odd raw moments,
whereas every even raw moment is fixed by a single endpoint-mixing
probability.  Second, an outcome-resolved tilted kernel propagates these local
sectors through the correlated record.  It yields exact boundary and memory
corrections at finite cycle number and, after a joint duration tilt, the
work-cumulant rate at fixed laboratory time.  The buffered-record Landauer
cost is retained as a compact thermodynamic consequence of the same memory
mode rather than as an independent theme.

The protocol is that of the companion experimental paper
\cite{companion}.  That work reports the apparatus, average energy ledger,
finite-time mean work, power, efficiency, and coherent--dephased performance
curves.  No experimental data or performance plots are repeated here.  The
mean-work identity is used as the entry point to the new quasiprobability and
fluctuation results.  Section~\ref{sec:model} defines the feedback cycle and
its outcome memory.  Section~\ref{sec:fcs} establishes the local interference
mechanism and its FCS signature.  Its propagation through correlated cycles
and laboratory time is developed in Sec.~\ref{sec:correlated}, while the
record cost and operational scope are summarized in Sec.~\ref{sec:reset}.
Detailed derivations and supporting figures are collected in the Appendices.

\section{Outcome-resolved feedback cycle and memory}
\label{sec:model}

\subsection{Feedback cycle}

\begin{figure}[tbp]
\centering

\resizebox{0.75\columnwidth}{!}{%
\begin{tikzpicture}[
x=1cm,
y=1cm,
>=Latex,
font=\footnotesize,
line width=0.65pt,
every node/.style={inner sep=1.5pt},
box/.style={
draw=black!65,
rounded corners=1.5pt,
minimum height=0.62cm,
align=center,
fill=white
},
plusbox/.style={
draw=blue!70!black,
rounded corners=1.5pt,
minimum height=0.62cm,
align=center,
fill=blue!5
},
minusbox/.style={
draw=red!70!black,
rounded corners=1.5pt,
minimum height=0.62cm,
align=center,
fill=red!4
},
recordbox/.style={
draw=orange!80!black,
rounded corners=1.5pt,
minimum height=0.62cm,
align=center,
fill=orange!7
},
mainarrow/.style={
-{Latex[length=2mm,width=1.3mm]},
draw=black!65
},
plusarrow/.style={
-{Latex[length=2mm,width=1.3mm]},
draw=blue!70!black
},
minusarrow/.style={
-{Latex[length=2mm,width=1.3mm]},
draw=red!70!black
}
]

% ============================================================
% 1. Initial state
% ============================================================

\node[
box,
minimum width=2.70cm
] (rin) at (0,0)
{
$\rho_{\mathrm{in}}^{(n)}$ at $H_0$
};

% ============================================================
% 2. Measurement
% ============================================================

\node[
box,
minimum width=2.50cm
] (meas) at (0,-1.15)
{
measurement $M_\theta$
};

\draw[mainarrow]
(rin.south)
--
(meas.north);

% ============================================================
% 3. Classical record
% ============================================================

\node[
recordbox,
minimum width=1.70cm
] (record) at (3.05,-1.15)
{
record\\[-1mm]
$X_n=a_n$
};

\draw[
-{Latex[length=2mm,width=1.3mm]},
draw=orange!80!black,
dashed
]
(meas.east)
--
node[
above,
text=orange!80!black,
font=\scriptsize
]
{write}
(record.west);

% ============================================================
% 4. Plus branch: Pi+
% ============================================================

\node[
plusbox,
minimum width=2.05cm
] (pip) at (-2.55,-2.45)
{
$\Pi_+$ at $H_0$
};

\draw[plusarrow]
(meas.south)
.. controls (-0.15,-1.80) and (-2.40,-1.67) ..
(pip.north);

\node[
text=blue!70!black,
font=\scriptsize
] at (-3.05,-1.82)
{
$a_n=+$
};

\node[
font=\scriptsize
] at (-3.80,-2.45)
{
$t_0^+$
};

% ============================================================
% 5. Compression process
%
% Pi+
%  |
% U_c ...
% tau_c ...
%  ↓
% rho_1,+
% ============================================================

% Pi+ 下方短竖线
\draw[
draw=blue!70!black,
line width=0.65pt
]
(pip.south)
--
(-2.55,-2.88);

% 过程文字
\node[
text=blue!70!black,
align=center,
font=\scriptsize
] (uc) at (-2.55,-3.22)
{
$U_c:\ H_0\rightarrow H_h$\\[0.55mm]
$\tau_c=t_1^+-t_0^+$
};

% rho_1,+ 节点
\node[
plusbox,
minimum width=2.20cm
] (rho1p) at (-2.55,-4.20)
{
$\rho_{1,+}$ at $H_h$
};

% 文字下方箭头
\draw[plusarrow]
(-2.55,-3.58)
--
(rho1p.north);

\node[
font=\scriptsize
] at (-3.88,-4.20)
{
$t_1^+$
};

% ============================================================
% 6. Expansion process
%
% rho_1,+
%  |
% U_e ...
% tau_e ...
%  ↓
% rho+
% ============================================================

% rho_1,+ 下方短竖线
\draw[
draw=blue!70!black,
line width=0.65pt
]
(rho1p.south)
--
(-2.55,-4.65);

% 过程文字
\node[
text=blue!70!black,
align=center,
font=\scriptsize
] (ue) at (-2.55,-5.02)
{
$U_e:\ H_h\rightarrow H_0$\\[0.55mm]
$\tau_e=t_2^+-t_1^+$
};

% rho+ 节点
\node[
plusbox,
minimum width=2.05cm
] (rhop) at (-2.55,-6.00)
{
$\rho_+$ at $H_0$
};

% 文字下方箭头
\draw[plusarrow]
(-2.55,-5.38)
--
(rhop.north);

\node[
font=\scriptsize
] at (-3.80,-6.00)
{
$t_2^+$
};

% ============================================================
% 7. Minus branch: Pi-
% ============================================================

\node[
minusbox,
minimum width=2.05cm
] (pim) at (2.55,-2.45)
{
$\Pi_-$ at $H_0$
};

\draw[minusarrow]
(meas.south)
.. controls (0.15,-1.80) and (2.40,-1.67) ..
(pim.north);

\node[
text=red!70!black,
font=\scriptsize
] at (3.05,-1.82)
{
$a_n=-$
};

\node[
font=\scriptsize
] at (3.80,-2.45)
{
$t_0^-$
};

% ============================================================
% 8. Cold-contact process
%
% Pi-
%  |
% Phi_- ...
% tau_d ...
%  ↓
% rho-
% ============================================================

% Pi- 下方短竖线
\draw[
draw=red!70!black,
line width=0.65pt
]
(pim.south)
--
(2.55,-3.59);

% 冷接触文字
\node[
text=red!70!black,
align=center,
font=\scriptsize
] (cold) at (2.55,-4.00)
{
$\Phi_-:$ cold contact at $H_0$\\[0.55mm]
$\tau_d=t_1^- - t_0^-$
};

% rho- 节点
\node[
minusbox,
minimum width=2.05cm
] (rhom) at (2.55,-6.00)
{
$\rho_-$ at $H_0$
};

% 文字下方箭头
\draw[minusarrow]
(2.55,-4.38)
--
(rhom.north);

\node[
font=\scriptsize
] at (3.80,-6.00)
{
$t_1^-$
};

% ============================================================
% 9. Next-cycle input state
% ============================================================

\node[
box,
minimum width=3.20cm
] (next) at (0,-7.38)
{
$\rho_{\mathrm{in}}^{(n+1)}
=
\rho_{a_n}$ at $H_0$
};

% ============================================================
% 10. Plus branch rejoins
% ============================================================

\draw[plusarrow]
(rhop.south)
.. controls (-2.55,-6.76) and (-1.25,-7.10) ..
(next.west);

% ============================================================
% 11. Minus branch rejoins
% ============================================================

\draw[minusarrow]
(rhom.south)
.. controls (2.55,-6.76) and (1.25,-7.10) ..
(next.east);

% ============================================================
% 12. Repeat
% ============================================================

\draw[
-{Latex[length=2mm,width=1.3mm]},
draw=black!65,
dotted
]
(next.south)
--
(0,-8.08);

\node[
font=\scriptsize
] at (0,-8.27)
{
repeat
};

\end{tikzpicture}%
}

\caption{
Outcome-resolved feedback protocol and time convention.  The
 measurement of $M_\theta$ prepares $\Pi_{a_n}$ and writes $a_n=\pm$ to the
 controller register $X_n$.  The $+$ outcome selects compression from $H_0$
 to $H_h$ followed by expansion back to $H_0$, with
 $U_{\rm ce}=U_eU_c$.  The $-$ outcome selects the cold-contact channel at
 fixed $H_0$.  The branch outputs become the inputs of the next cycle.
 Reset of $X_n$ requires a separate physical specification and is not a
 working-medium stroke.}
 \label{fig:protocol}
\end{figure}
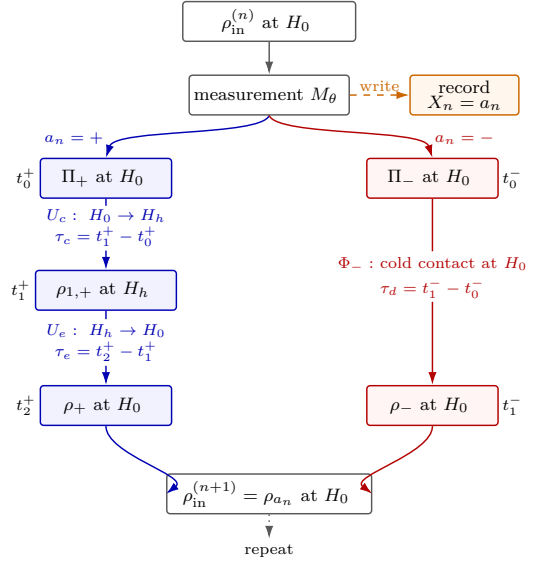

At the beginning of cycle $n$, the working medium has state
$\rho_{\rm in}^{(n)}$ and Hamiltonian
\begin{equation}
 H_0=\frac{\hbar\omega_c}{2}\sigma_z.
 \label{eq:H0}
\end{equation}
An instantaneous projective measurement of
\begin{equation}
 M_\theta=\cos\theta\,\sigma_z+\sin\theta\,\sigma_x
 \label{eq:measurement-observable}
\end{equation}
has projectors $\Pi_a=(\id+aM_\theta)/2$, where $a=\pm1$ labels the two
outcomes $+$ and $-$ \cite{WangXue2022}.  The outcome $a_n$ is written to the classical
controller register $X_n$ \cite{Cottet2017,Masuyama2018} and selects one of two feedback branches \cite{Yan2024}, as shown
in Fig.~\ref{fig:protocol}.

On the $+$ branch, $t_0^+$, $t_1^+$, and $t_2^+$ denote the
postmeasurement time, the end of compression, and the end of expansion,
respectively.  The Hamiltonian follows $H_0\to H_h\to H_0$, where
$H_h=\hbar\omega_h\sigma_x/2$.  The stroke durations are
$\tau_c=t_1^+-t_0^+$ and $\tau_e=t_2^+-t_1^+$, and the corresponding
propagators are $U_c=U(t_1^+,t_0^+)$ and $U_e=U(t_2^+,t_1^+)$.  We denote
their composite compression--expansion propagator by
$U_{\rm ce}=U_eU_c$.  On the $-$ branch, the Hamiltonian remains $H_0$ and
the cold-contact channel $\Phi_-$ acts for
$\tau_d=t_1^--t_0^-$.  The conditional output states are therefore
\begin{align}
 \rho_{1,+}&=U_c\Pi_+U_c^\dagger,
 &\rho_+&=U_e\rho_{1,+}U_e^\dagger
          =U_{\rm ce}\Pi_+U_{\rm ce}^\dagger,\nonumber\\
 \rho_-&=\Phi_-(\Pi_-).
 \label{eq:conditional-output}
\end{align}
For the theory curves, the unitary strokes follow the same interpolation as
the companion protocol \cite{companion}.  With a local stroke
coordinate $s\in[0,1]$,
\begin{align}
 \Omega_c(s)&=\omega_c(1-s)+\omega_hs,\nonumber\\
 \Omega_e(s)&=\omega_h(1-s)+\omega_cs,\nonumber\\
 H_c(s)&=\frac{\hbar\Omega_c(s)}{2}
 \left[\cos\!\left(\frac{\pi s}{2}\right)\sigma_z
 +\sin\!\left(\frac{\pi s}{2}\right)\sigma_x\right],\nonumber\\
 H_e(s)&=\frac{\hbar\Omega_e(s)}{2}
 \left[\cos\!\left(\frac{\pi s}{2}\right)\sigma_x
 +\sin\!\left(\frac{\pi s}{2}\right)\sigma_z\right].
 \label{eq:driving-paths}
\end{align}
Thus $H_c(0)=H_e(1)=H_0$, $H_c(1)=H_e(0)=H_h$, and
\begin{equation}
 U_\alpha=\mathcal T_{\leftarrow}\exp\!\left[
 -\frac{\ii\tau_\alpha}{\hbar}\int_0^1H_\alpha(s)\,ds\right],
 \qquad \alpha=c,e.
 \label{eq:numerical-propagators}
\end{equation}

In the ordered energy basis \((\lvert e\rangle,\lvert g\rangle)\) of \(H_0\), the excited and ground states satisfy \(H_0\lvert e\rangle=(\hbar\omega_c/2)\lvert e\rangle\) and \(H_0\lvert g\rangle=-(\hbar\omega_c/2)\lvert g\rangle\), respectively, and \(\rho_{eg}\equiv\langle e\rvert\rho\lvert g\rangle\). The ideal cold-contact channel is specified in this basis by
\begin{align}
 z(\tau_d)&=z_\beta+[z(0)-z_\beta]\ee^{-\Gamma\tau_d},\nonumber\\
 \rho_{eg}(\tau_d)&=\rho_{eg}(0)
 \ee^{-(\Gamma/2+\ii\omega_c)\tau_d},
 \label{eq:compact-bath}
\end{align}
where \(z=\operatorname{Tr}(\rho\sigma_z)\), \(\beta=(k_{\mathrm B}T_c)^{-1}\) is the inverse temperature of the cold bath, \(\gamma\) is the bare population-relaxation rate, \(n_{\rm th}=(e^{\beta\hbar\omega_c}-1)^{-1}\), \(z_\beta=-\tanh(\beta\hbar\omega_c/2)\), and \(\Gamma=\gamma(2n_{\rm th}+1)\) is the total population-relaxation rate. 
Unless stated otherwise, the parameters are
$\omega_c=20\pi$, $\omega_h=30\pi$, $\beta\epsc=1$, $\gamma=1$,
$\hbar=1$, $\tau_c=0.1$, and $\tau_e=0.01$.

The ideal map in Eq.~\eqref{eq:compact-bath} is used for the numerical
illustrations.  The analytical constructions below require only the
branch-conditioned completely positive trace-preserving maps and therefore
do not depend on its specific transverse-dephasing rate.  The numerical
transition probabilities and the subleading eigenvalue $\lambda_2$ governing
the outcome memory, however, depend quantitatively on the chosen map.

The realized output becomes the input of the next cycle,
$\rho_{\rm in}^{(n+1)}=\rho_{a_n}$.  Because the two branches generally
have different durations, $t_2^+$ and $t_1^-$ are distinct physical
endpoints.  This distinction is retained in the laboratory-time statistics
below.  The intermediate $H_h$ contributions cancel when the compression and
expansion mean-work increments are added, so their sum equals the endpoint
energy change generated by $U_{\rm ce}$.  A controller-memory reset, if
implemented, is a separately specified operation rather than an additional
working-medium stroke \cite{Strasberg2017}. The experimental realization and full dynamical generators are given in
Ref.~\cite{companion}; no experimental data are replotted here.

The probability of obtaining outcome $b$ in the next cycle, conditioned on
outcome $a$ in the present cycle, is
\begin{equation}
 q_{b|a}
 =
 \Tr(\Pi_b\rho_a).
 \label{eq:conditional-probability}
\end{equation}
Because each rank-one projective outcome $a$ prepares the fixed state
$\Pi_a$, independently of the incoming state, $q_{b|a}$ depends only on the
current outcome.  The outcome record therefore forms an exact
time-homogeneous first-order Markov chain.  With column probability vectors,
its evolution is
\begin{equation}
 \bm p^{(n+1)}=R\bm p^{(n)},\qquad R_{ba}=q_{b|a}.
 \label{eq:general-outcome-map}
\end{equation}
The columns of $R$ sum to unity, and a stationary distribution $\bm p^*$
satisfies $R\bm p^*=\bm p^*$.

\subsection{Outcome memory}

For a real-valued function $f_a$ of a stationary outcome $A_n$, define
$\widetilde f_a=f_a-\sum_b p_b^*f_b$.  Stationarity gives
$\Pr(A_n=a,A_{n+\ell}=b)=[R^\ell]_{ba}p_a^*$,
{where $\Pr(A_n=a,A_{n+\ell}=b)$ denotes the joint probability
of observing outcomes $a$ and $b$ separated by $\ell$ cycles. The corresponding stationary lag-$\ell$ covariance
of the observable $f$ is then}
\begin{equation}
 C_f(\ell)
 =
 \left\langle
 \widetilde f_{A_{n+\ell}}\widetilde f_{A_n}
 \right\rangle_*
 =
 \widetilde{\bm f}^{\mathsf T}R^\ell
 \operatorname{diag}(\bm p^*)\widetilde{\bm f}.
 \label{eq:general-covariance}
\end{equation}
Only nonstationary modes of $R$ contribute after centering.  Consequently,
the stationary composition $\bm p^*$ does not determine the temporal
correlations \cite{Norris1997}.

For the present two-outcome engine, write
\begin{equation}
R=
\begin{pmatrix}
1-r&s\\
r&1-s
\end{pmatrix},
\qquad
r=q_{-|+},\quad s=q_{+|-}.
\label{eq:binary-outcome-map}
\end{equation}
For $r+s>0$, the stationary probabilities are
\begin{equation}
 p_+^*=\frac{s}{r+s},
 \qquad
 p_-^*=\frac{r}{r+s}.
 \label{eq:stationary}
\end{equation}
The nonstationary eigenvalue is
\begin{equation}
 \lambda_2=1-r-s.
 \label{eq:lambda2}
\end{equation}
For an irreducible and aperiodic record, $|\lambda_2|<1$.  This eigenvalue is
the one-cycle multiplier of the sole nonstationary mode and therefore carries
all binary outcome memory.  It is independent of the stationary weight in the
sense that fixed $p_+^*$ fixes the ratio $s/r$, whereas $\lambda_2$ depends on
their sum.

Let $I_n=1$ when the outcome in cycle $n$ is $+$ and $I_n=0$ otherwise.
Since $[R^\ell]_{++}=p_+^*+p_-^*\lambda_2^\ell$, stationarity gives
$\Pr(I_n=1,I_{n+\ell}=1)=p_+^*[R^\ell]_{++}$ and hence
\begin{equation}
C_I(\ell)
\equiv
\operatorname{Cov}(I_n,I_{n+\ell})
=p_+^*p_-^*\lambda_2^\ell.
\label{eq:indicator-covariance}
\end{equation}
Thus $C_I(\ell)/C_I(0)=\lambda_2^\ell$ whenever
$p_+^*p_-^*>0$.  Positive $\lambda_2$ produces persistent correlations,
negative $\lambda_2$ produces alternating correlations, {
and $\lambda_2=0$ gives independent measurement
outcomes, although adjacent work increments
need not be independent.
}

\section{Coherence--transition interference in a single cycle}
\label{sec:fcs}

The \(+\) outcome both selects the driven feedback branch and prepares the
state on which the drive acts.  Work is taken as positive when it is done on
the qubit, so extracted work has the opposite sign.  Throughout this paper,
the counted work is the work done on the qubit during the driven
compression--expansion branch.  The measurement-induced energy change and
the heat exchanged during the cold-contact branch enter separate energy
ledgers, as in the companion paper, but are not
included in the work counting field.  Accordingly, the \(-\) branch carries
zero counted work.  The mean-work decomposition from that paper is retained
only to identify the microscopic quantity whose FCS signature is resolved
below.

\subsection{Population and coherence contributions}

In the ordered energy basis $(|e\rangle,|g\rangle)$, the state prepared by the
$+$ outcome is
\begin{equation}
 \Pi_+=\frac12
 \begin{pmatrix}
  1+\cos\theta&\sin\theta\\
  \sin\theta&1-\cos\theta
 \end{pmatrix}.
 \label{eq:measurement-state-energy-basis}
\end{equation}
Thus the measurement angle reallocates backaction between a population bias,
$\rho_{ee}-\rho_{gg}=\cos\theta$, and energy coherence,
$C_{\ell_1}(\Pi_+)=|\sin\theta|$ \cite{Baumgratz2014,Streltsov2017}.

Let $U_{\rm ce}=U_eU_c$ be the complete finite-time
compression--expansion propagator defined in
Sec.~\ref{sec:model}. Its residual endpoint mixing is
\begin{equation}
 \mathcal T_{\rm ce}
 \equiv |(U_{\rm ce})_{eg}|^2
 =|(U_{\rm ce})_{ge}|^2 .
 \label{eq:nonadiabatic-mixing}
\end{equation}
The state $\Pi_+$ entering the driven branch is active and
coherent rather than Gibbsian \cite{Korzekwa2016,Lostaglio2015}.

To isolate the effect of the initial energy coherence, consider
an arbitrary qubit state $\rho$. Let
$\Delta_0[\rho]=\sum_{n=e,g}\rho_{nn}|n\rangle\langle n|$
denote complete dephasing in the eigenbasis of $H_0$, and let
$p_e^f(\rho)=
\langle e|U_{\rm ce}\rho U_{\rm ce}^\dagger|e\rangle$
be the final excited-state probability. The dimensionless
coherence--transition interference contribution is
\begin{align}
 \nu_{\rm FC}[\rho]
 &\equiv
 p_e^f(\rho)-p_e^f\!\left(\Delta_0[\rho]\right)
 \nonumber\\
 &=
 2\operatorname{Re}\!\left[
 (U_{\rm ce})_{eg}(U_{\rm ce})_{ee}^*\rho_{ge}
 \right].
 \label{eq:measurement-interference-general}
\end{align}
Thus $\nu_{\rm FC}[\rho]$ is the signed change in the final
excited-state probability caused by retaining the initial
energy coherence while keeping the initial populations fixed.
It originates from interference between the amplitudes
$|e\rangle\rightarrow|e\rangle$ and
$|g\rangle\rightarrow|e\rangle$ and applies to both pure and
mixed initial states.

For the measurement-prepared state, we write
$\nu_{\rm FC}\equiv\nu_{\rm FC}[\Pi_+]$. Its exact conditional
mean work is
\begin{equation}
 \bar w_+=w_{\rm pop}+w_{\rm coh},
 \label{eq:work-mechanism}
\end{equation}
where
$w_{\rm pop}=-\epsc\mathcal T_{\rm ce}\cos\theta$
is the population-transfer contribution and
$w_{\rm coh}=\epsc\nu_{\rm FC}$
is the coherence--transition interference contribution.

Here work done on the qubit is taken as positive, whereas the companion
experimental paper \cite{companion} takes extracted work as positive.
Including
the stationary probability of selecting the driven branch
therefore gives
\begin{equation}
 \langle w\rangle_{\rm ext}
 =
 -p_+^*\bar w_+
 =
 p_+^*\epsc
 \left(
 \mathcal T_{\rm ce}\cos\theta-\nu_{\rm FC}
 \right).
 \label{eq:extracted-mean-work}
\end{equation}
To make contact with the stroke-resolved notation of the companion paper, we recall the nonadiabatic transition probabilities introduced there $\xi_c=|\langle e(t_1^+)|U_c|g(t_0^+)\rangle|^2$ and $\xi_e=|\langle e(t_2^+)|U_e|g(t_1^+)\rangle|^2$, together with the corresponding coherence-dependent dynamical terms
$\zeta_c=\operatorname{Re}[
\langle g(t_1^+)|U_c|g(t_0^+)\rangle
\allowbreak
\langle g(t_0^+)|\Pi_+|e(t_0^+)\rangle
\allowbreak
\langle e(t_0^+)|U_c^\dagger|g(t_1^+)\rangle]$
and
$\zeta_e=\operatorname{Re}[
\langle g(t_2^+)|U_e|g(t_1^+)\rangle
\allowbreak
\langle g(t_1^+)|\rho_{1,+}|e(t_1^+)\rangle
\allowbreak
\langle e(t_1^+)|U_e^\dagger|g(t_2^+)\rangle]$.
Here $|g(t)\rangle$ and $|e(t)\rangle$ denote the instantaneous ground and
excited eigenstates of the driven Hamiltonian at the indicated endpoints.
With these definitions, Eq.~\eqref{eq:extracted-mean-work} is algebraically
identical to the stroke-resolved mean-work expression of the companion paper,
because
\begin{align}
 \mathcal T_{\rm ce}\cos\theta-\nu_{\rm FC}
 &=
 2\left(
 \zeta_c+\zeta_e-2\xi_e\zeta_c
 \right)
 \nonumber\\
 &\quad+
 \left(
 \xi_c+\xi_e-2\xi_c\xi_e
 \right)\cos\theta .
 \label{eq:bridge-stroke-mechanism}
\end{align}

When the product in
Eq.~\eqref{eq:measurement-interference-general} is nonzero,
define
\[
 \phi_{\rm rel}
 =
 \arg\!\left[
 (U_{\rm ce})_{eg}(U_{\rm ce})_{ee}^*\rho_{ge}
 \right].
\]
For $\rho=\Pi_+$ and $0\leq\theta\leq\pi/2$, this gives
\begin{equation}
 \nu_{\rm FC}
 =
 \sin\theta
 \sqrt{\mathcal T_{\rm ce}(1-\mathcal T_{\rm ce})}
 \cos\phi_{\rm rel}.
 \label{eq:measurement-interference}
\end{equation}
If the product vanishes, $\nu_{\rm FC}=0$ directly and no
relative phase is required.

The conventional meaning of $\mathcal T_{\rm ce}$ is fixed by
comparison with the passive Gibbs state
$\rho_\beta=\ee^{-\beta H_0}/\Tr(\ee^{-\beta H_0})$:
\begin{align}
 W_{\rm if}^{(\beta)}
 &=
 \epsc\mathcal T_{\rm ce}\tanh(\beta\epsc/2)
 \nonumber\\
 &=
 \beta^{-1}
 D\!\left(
 U_{\rm ce}\rho_\beta U_{\rm ce}^\dagger
 \Vert\rho_\beta
 \right)
 \geq0,
 \label{eq:inner-friction-reference}
\end{align}
where
$D(\rho\Vert\sigma)=
\Tr[\rho(\ln\rho-\ln\sigma)]$. {The quantity
$W_{\mathrm{if}}^{(\beta)}$ represents the inner-friction
energy cost of the complete driven branch evaluated for
the passive Gibbs reference state $\rho_\beta$. Equivalently,
$W_{\mathrm{if}}^{(\beta)}
=
\operatorname{Tr}\!\left[
H_0\left(
U_{ce}\rho_\beta U_{ce}^{\dagger}-\rho_\beta
\right)
\right]$.
Since the initial and final Hamiltonians coincide with
$H_0$, the adiabatic reference process leaves the Gibbs
state unchanged. Thus, $W_{\mathrm{if}}^{(\beta)}$
quantifies the excess endpoint energy due to nonadiabatic
transitions for a passive Gibbs input, consistent with
the conventional interpretation of quantum inner friction
\cite{Insinga2020,Lee2020}. Its nonnegativity follows
directly from Eq.~\eqref{eq:inner-friction-reference}
and the positivity of quantum relative entropy. It should
be distinguished from the conditional mean work
$\bar{w}_+$ of the measurement-prepared state $\Pi_+$,
which contains both population-transfer and
coherence--transition interference contributions.
}
Equations~\eqref{eq:measurement-interference-general},
\eqref{eq:work-mechanism}, and
\eqref{eq:inner-friction-reference} are derived in
Appendix~\ref{app:fcs}.

Using the stroke-resolved transition probabilities $\xi_c$ and $\xi_e$
defined above, the cycle-averaged quantum-inner-friction contribution to the
extracted work used in the companion paper is
\(
 \langle w\rangle_{\rm fri}
 =
 p_+^*\epsc\cos\theta
 \left(
 \xi_c+\xi_e-2\xi_c\xi_e
 \right)
 =
 p_+^*\epsc\cos\theta
 \left[
 \xi_c(1-\xi_e)+\xi_e(1-\xi_c)
 \right].
 \)
Since $0\leq\xi_c,\xi_e\leq1$,
$\langle w\rangle_{\rm fri}\geq0$ for
$0\leq\theta\leq\pi/2$. It is the quantum-inner-friction
part of the total extracted work
$\langle w\rangle_{\rm ext}$, rather than the total work
itself. With the sign convention of the present Article, its
contribution to work done on the qubit is
$-\langle w\rangle_{\rm fri}$.

The quantities $\langle w\rangle_{\rm fri}$,
$W_{\rm if}^{(\beta)}$, and $-p_+^*w_{\rm pop}$ are not
identical. The first is the stroke-resolved contribution used in the companion
 paper, the second is the passive-reference
inner-friction cost of the complete propagator, and the third
is the endpoint population-transfer contribution in the
present decomposition. These descriptions are complementary
and should not be identified term by term.

Accordingly, $\mathcal T_{\rm ce}$ measures the residual
endpoint mixing of the complete driven branch and is not a
sum of two independently positive stroke frictions. For
$0\leq\theta\leq\pi/2$, the measurement-prepared state is
excited-state biased and $w_{\rm pop}\leq0$, so finite-time
population transfer can release energy. The coherent
contribution can reinforce or oppose this extraction and
vanishes without initial energy coherence, for
$\mathcal T_{\rm ce}=0$ or $1$, or for
$\cos\phi_{\rm rel}=0$. Coherence magnitude alone therefore
does not determine engine performance \cite{Xu2018,LuJiangRen2024}.

\subsection{FCS signature of the interference}

Because the $+$ branch begins at $t_0^+$ and ends at $t_2^+$ with
$H(t_0^+)=H(t_2^+)=H_0$, its symmetric FCS quasicharacteristic function for
$\rho=\Pi_+$ is \cite{Solinas2015}
\begin{equation}
 \chi_+(u)=\Tr\!\left[
 \ee^{\ii uH_0/2}U_{\rm ce}\ee^{-\ii uH_0/2}\Pi_+
 \ee^{-\ii uH_0/2}U_{\rm ce}^\dagger\ee^{\ii uH_0/2}
 \right].
 \label{eq:single-characteristic}
\end{equation}
No energy projection or counting operation is inserted at the junction
$t_1^+$; Eq.~\eqref{eq:single-characteristic} counts the net endpoint work
of the combined unitary branch.
It obeys $\chi_+(0)=1$ and $\chi_+(-u)=\chi_+(u)^*$, but its Fourier
transform can be signed because $[\Pi_+,H_0]\ne0$ for generic $\theta$.
This ordering is distinct
from both an initial projective energy measurement and the $q=0$
Margenau--Hill construction \cite{Pei2023}.  

For the present qubit,
Eq.~\eqref{eq:single-characteristic} reduces to
\begin{align}
 \chi_+(u)&=1-\mathcal T_{\rm ce}
 +\mathcal T_{\rm ce}\cos(u\epsc)
 -\ii\mathcal T_{\rm ce}\cos\theta\sin(u\epsc)\nonumber\\
 &\quad+2\ii\nu_{\rm FC}\sin(u\epsc/2).
 \label{eq:single-energy-expansion}
\end{align}
The integer-frequency terms describe population transfer, whereas the
half-frequency term is the coherence--transition interference.  Midpoint
work values are a general feature of symmetric FCS with initial energy
coherence.  For the present equal-endpoint qubit branch, however, the entire
signed sector is one equal-and-opposite half-quantum pair whose magnitude is
the coherent contribution to the mean work.  The quasiprobability has support
at $w/\epsc\in\{-1,-1/2,0,1/2,1\}$.

\begin{figure}[h]
\vspace{0.15cm}
\hspace*{0.3cm}
\begin{overpic}[width=3.4cm]
{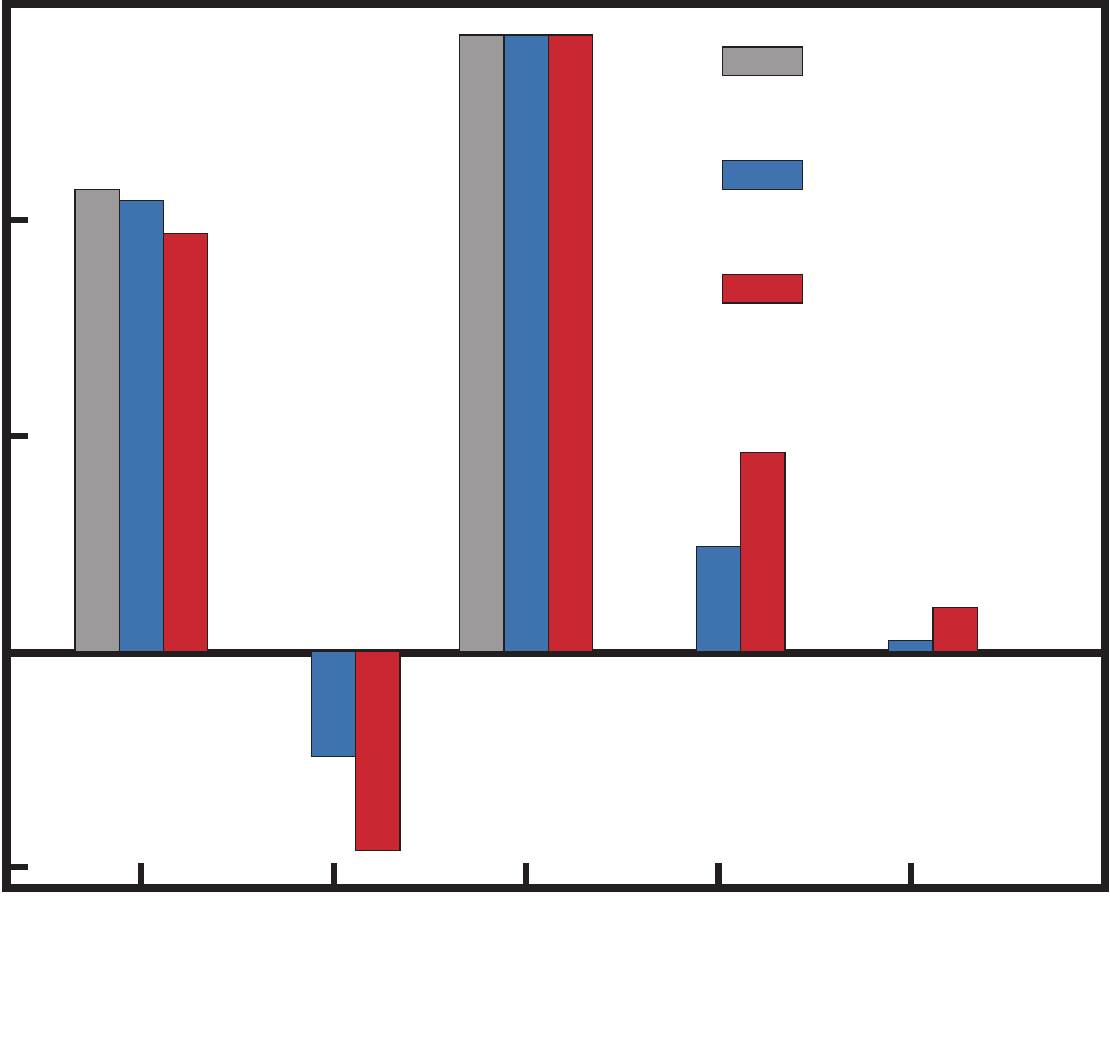}
\put(0,100){\scalebox{0.9}{(a)}}
\put(-20,45){\rotatebox{90}{\scalebox{0.8}{$\mathcal P_{\rm FC}(w|+)$}}}
\put(44,-4){\scalebox{0.8}{$w/(\hbar\omega_c)$}}
\put(-10,93){\scalebox{0.7}{0.6}}
\put(-10,74){\scalebox{0.7}{0.4}}
\put(-10,55){\scalebox{0.7}{0.2}}
\put(-7,35){\scalebox{0.7}{0}}
\put(-12,16){\scalebox{0.7}{-0.2}}
\put(10,6.5){\scalebox{0.7}{-1}}
\put(25,6.5){\scalebox{0.7}{$-\frac{1}{2}$}}
\put(46,6.5){\scalebox{0.7}{0}}
\put(63,6.5){\scalebox{0.7}{$\frac{1}{2}$}}
\put(80,6.5){\scalebox{0.7}{1}}
\put(74,88){\scalebox{0.7}{$\theta=0$}}
\put(74,78){\scalebox{0.7}{$\theta=0.1\pi$}}
\put(74,68){\scalebox{0.7}{$\theta=0.2\pi$}}
\end{overpic}
\centering
\hspace{0.8cm} 
\vspace{0.15cm}
\begin{overpic}[width=3.4cm]
{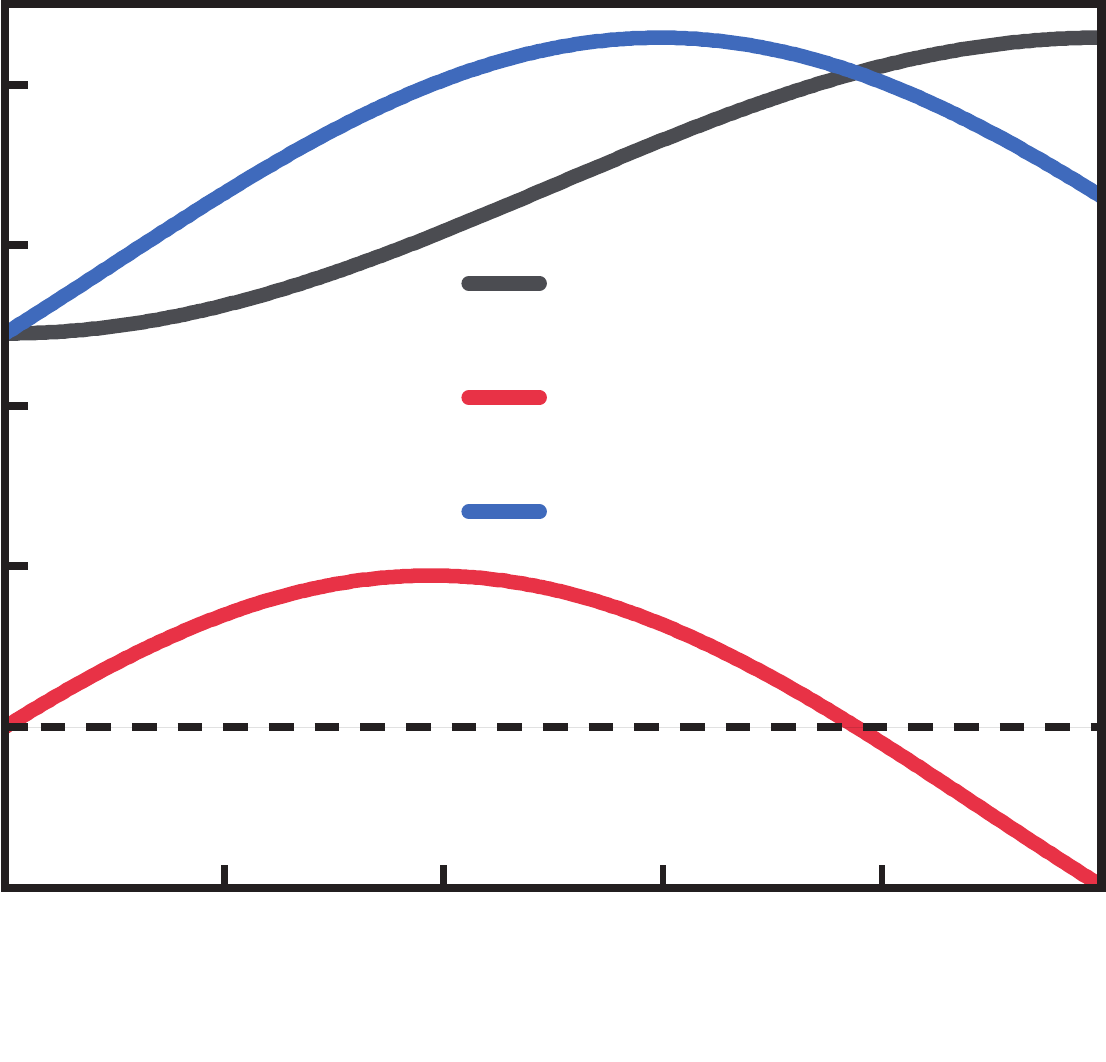}
\put(0,100){\scalebox{0.9}{(b)}}
\put(44,-6){\scalebox{1}{$\theta/\pi$}}
\put(52,69){\scalebox{0.7}{$\kappa_{2,+}^{\Delta}$}}
\put(52,59){\scalebox{0.7}{$\Delta_{\rm coh}\kappa_{2,+}$}}
\put(52,48){\scalebox{0.7}{$\kappa_{2,+}$}}
\put(0,7){\scalebox{0.7}{0}}
\put(15,7){\scalebox{0.7}{0.1}}
\put(35,7){\scalebox{0.7}{0.2}}
\put(55,7){\scalebox{0.7}{0.3}}
\put(75,7){\scalebox{0.7}{0.4}}
\put(95,7){\scalebox{0.7}{0.5}}
\put(-12,14){\scalebox{0.7}{-0.1}}
\put(-7,28){\scalebox{0.7}{0}}
\put(-10,43){\scalebox{0.7}{0.1}}
\put(-10,57){\scalebox{0.7}{0.2}}
\put(-10,72){\scalebox{0.7}{0.3}}
\put(-10,86){\scalebox{0.7}{0.4}}
\end{overpic}
\addtolength{\abovecaptionskip}{-3pt}
\caption{Mechanism-resolved symmetric FCS statistics for the unitary branch.
 (a) The five weights at three measurement angles.  Integer-quantum peaks are
 produced by population transfer through the nonadiabatic channel; the
 equal-and-opposite half-quantum pair is the measurement-coherence
 interference.  (b) Dephased second-cumulant baseline
 $\kappa_{2,+}^{\Delta}$, signed coherent correction
 $\Delta_{\rm coh}\kappa_{2,+}$, and their sum.  The sign change of the
 coherent correction demonstrates that coherence can either enhance or
 suppress work fluctuations.  Parameters are $\tau_c=0.1$,
 $\tau_e=0.01$, $\omega_c=20\pi$, $\omega_h=30\pi$, and $\hbar=1$.}
\label{fig:work-quasiprobability}
\vspace{0cm}
\end{figure}
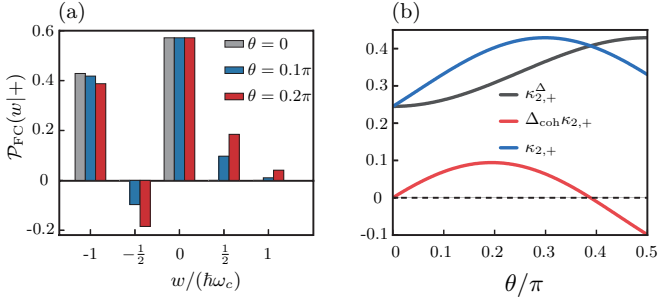

Although $H_h$ affects $U_{\rm ce}$ and therefore the weights, it is not a
counted endpoint; hence $\omega_h$ does not enlarge this five-point support.
The model-specific weights are
\begin{subequations}
\label{eq:five-weights}
\begin{align}
 \mathcal P_{\rm FC}(+\epsc|+)
 &=\frac{\mathcal T_{\rm ce}}{2}(1-\cos\theta),
 \label{eq:weight-plus}\\
 \mathcal P_{\rm FC}(-\epsc|+)
 &=\frac{\mathcal T_{\rm ce}}{2}(1+\cos\theta),
 \label{eq:weight-minus}\\
 \mathcal P_{\rm FC}(0|+)&=1-\mathcal T_{\rm ce},
 \label{eq:weight-zero}\\
 \mathcal P_{\rm FC}(+\epsc/2|+)&=\nu_{\rm FC},
 \label{eq:weight-half-plus}\\
 \mathcal P_{\rm FC}(-\epsc/2|+)&=-\nu_{\rm FC}.
 \label{eq:weight-half-minus}
\end{align}
\end{subequations}
The integer sector is nonnegative and normalized by itself.  The midpoint
sector has zero total weight and directly exposes the interference that is
lost after energy dephasing.  {
To quantify the negativity of the symmetric-FCS work
quasiprobability, we define the full-counting-statistics
(FCS) negativity $\mathcal N_{\mathrm{FC}}$ as one-half of the
excess of the $L^1$ norm of the quasiprobability weights
over unity:}
\begin{equation}
 \mathcal N_{\rm FC}=\frac12\left[
 \sum_j|\mathcal P_{\rm FC}(w_j|+)|-1\right].
 \label{eq:negativity}
\end{equation}
{For the present five-point distribution, this gives}
\begin{equation}
 \mathcal N_{\rm FC}=|\nu_{\rm FC}|=|w_{\rm coh}|/\epsc.
 \label{eq:negativity-identity}
\end{equation}
This identity is specific to the symmetric ordering and the present
equal-endpoint qubit branch.

The full moment hierarchy makes the physical division transparent:
\begin{align}
 \langle w^{2k}\rangle_+&=(\epsc) ^{2k}\mathcal T_{\rm ce},
 &&k\ge1,\nonumber\\
 \langle w^{2k+1}\rangle_+
 &=(\epsc)^{2k+1}[-\mathcal T_{\rm ce}\cos\theta+4^{-k}\nu_{\rm FC}],
 &&k\ge0.
 \label{eq:moment-parity}
\end{align}
The detailed derivations of the five quasiprobability weights in
Eq.~\eqref{eq:five-weights}, the negativity relation in
Eq.~\eqref{eq:negativity-identity}, and the odd--even moment hierarchy in
Eq.~\eqref{eq:moment-parity} are given in
Appendix~\ref{app:fcs}.

For the unitary qubit branch, all even raw moments are
determined solely by the residual endpoint mixing,
whereas measurement coherence contributes to the odd
raw moments through the half-quantum sector.
 {
The corresponding second work cumulant,
$\kappa_{2,+}\equiv
\langle w^2\rangle_+-\langle w\rangle_+^2$,
characterizes the conditional work fluctuations.
Since $\langle w\rangle_+=\bar w_+$ and
$\langle w^2\rangle_+=(\epsc)^2\mathcal T_{\rm ce}$,}
it follows that
\begin{equation}
\kappa_{2,+}
=(\epsc)^2\mathcal T_{\rm ce}-\bar w_+^2\geq 0.
\label{eq:conditional-moments}
\end{equation}

To isolate the coherence contribution to the work fluctuations, let
\(\kappa_{2,+}^{\Delta}\) denote the second work cumulant obtained after
replacing the initial state \(\Pi_+\) by its energy-dephased counterpart
\(\Delta_0[\Pi_+]\), which has the same energy populations.  For this
energy-diagonal state, \(\kappa_{2,+}^{\Delta}\) is the population-only work
variance.  Normalizing by \((\epsc)^2\) gives
\begin{align}
 \frac{\kappa_{2,+}^{\Delta}}{(\epsc)^2}
 &=\mathcal T_{\rm ce}
   -\mathcal T_{\rm ce}^2\cos^2\theta,\nonumber\\
 \frac{\Delta_{\rm coh}\kappa_{2,+}}{(\epsc)^2}
 &\equiv
 \frac{\kappa_{2,+}-\kappa_{2,+}^{\Delta}}{(\epsc)^2}
 =2\mathcal T_{\rm ce}\cos\theta\,\nu_{\rm FC}
  -\nu_{\rm FC}^2.
 \label{eq:coherence-cumulant-correction}
\end{align}
Thus \(\Delta_{\rm coh}\kappa_{2,+}\) is the signed change in the conditional
work variance caused by retaining the initial energy coherence.  Its sign is
determined by
\(\nu_{\rm FC}(2\mathcal T_{\rm ce}\cos\theta-\nu_{\rm FC})\), rather than by
\(C_{\ell_1}(\Pi_+)\) or \(\mathcal N_{\rm FC}\) alone.  Consequently,
coherence magnitude and FCS negativity do not by themselves determine whether
the conditional work fluctuations are enhanced or suppressed.

Figure~\ref{fig:work-quasiprobability}(a) resolves the five-point
quasiprobability into its population and coherence-interference sectors.
At \(\theta=0\), the state \(\Pi_+\) is energy diagonal, so the half-quantum
weights vanish and only the classical values
\(w/\epsc=-1,0,+1\) remain.  As \(\theta\) increases, the excited-state bias
is reduced: weight is transferred from the negative-work peak at
\(-\epsc\) to the positive-work peak at \(+\epsc\), while the zero-work
weight \(1-\mathcal T_{\rm ce}\) remains unchanged.  At the same time,
measurement coherence generates the equal-and-opposite pair at
\(w=\pm\epsc/2\).  For the parameters shown, the negative weight occurs at
\(-\epsc/2\), directly revealing the nonclassical FCS sector.  Because the
two half-quantum weights have opposite signs, they cancel in the
normalization and in every even raw moment, but add in the odd moments,
including the coherent mean-work contribution
\(w_{\rm coh}=\epsc\nu_{\rm FC}\).

Figure~\ref{fig:work-quasiprobability}(b) shows how this interference changes
the conditional work fluctuations.  The black curve is the population-only
baseline \(\kappa_{2,+}^{\Delta}\) obtained after energy dephasing,
the red curve is the signed coherence correction
\(\Delta_{\rm coh}\kappa_{2,+}\), and the blue curve is their sum,
\(\kappa_{2,+}\).  At small and intermediate \(\theta\), the term
\(2\mathcal T_{\rm ce}\cos\theta\,\nu_{\rm FC}\) dominates, so coherence
enhances the work variance.  At larger \(\theta\), the negative term
\(-\nu_{\rm FC}^2\) becomes dominant; the correction changes sign and
coherence suppresses the variance.  In particular, at
\(\theta=\pi/2\) the linear term vanishes and
{\(\Delta_{\rm coh}\kappa_{2,+}=-(\epsc)^2\nu_{\rm FC}^2<0\)}.
The resulting maximum and subsequent decrease of the full second cumulant
show that increasing measurement coherence does not imply increasing work
fluctuations.  This sign-changing fluctuation effect is not contained in the
mean-work curves reported in the companion paper.

{
The five-point structure is specific to the
equal-endpoint qubit branch under symmetric FCS.
It explicitly separates population transfer
from phase-sensitive coherence interference,
yielding distinct contributions to odd and even
raw work moments.
However, the single-branch characteristic
function does not retain work--next-outcome
correlations, whose propagation across feedback
cycles requires the outcome-resolved tilted
kernel developed in Sec.~IV.
}

\section{Propagation through outcome memory}
\label{sec:correlated}

The FCS signature established in Sec.~\ref{sec:fcs} is conditional on a
single driven branch.  In stationary operation, the driven branch is selected
with probability $p_+^*$, whereas the thermal branch contributes zero driven
work.  The stationary one-cycle characteristic function is therefore
$\chi_1(u)=p_-^*+p_+^*\chi_+(u)$, and the mean work per completed cycle is
$\bar w=p_+^*\bar w_+$.  The second cumulant of this one-cycle marginal,
which also provides {the independent-work-increment baseline}, is
\begin{equation}
 \kappa_{2,1}
 =p_+^*{(\epsc)^2}\mathcal T_{\rm ce}-\bar w^2.
 \label{eq:one-cycle-cumulants}
\end{equation}
Equation~\eqref{eq:one-cycle-cumulants} includes the stationary mixing of the
driven and zero-work branches, but it contains no correlations between
different cycles.

\subsection{Accumulated work over completed cycles}
\label{sec:tilted}

The work produced in one cycle and the outcome that selects the following
cycle must be retained jointly.  For this purpose, define the tilted branch
maps
\begin{align}
 \mathcal K_+^{(u)}(\rho)
 &=
 \ee^{\ii uH_0/2}U_{\rm ce}\ee^{-\ii uH_0/2}\rho
 \ee^{-\ii uH_0/2}U_{\rm ce}^\dagger\ee^{\ii uH_0/2},
 \nonumber\\
 \mathcal K_-^{(u)}(\rho)&=\Phi_-(\rho),
 \label{eq:tilted-maps}
\end{align}
and the outcome-resolved tilted kernel
\begin{equation}
 [\mathsf R(u)]_{ba}
 =\Tr\!\left[\Pi_b\mathcal K_a^{(u)}(\Pi_a)\right].
 \label{eq:resolved-instrument}
\end{equation}
At $u=0$, the physical outcome map is recovered,
$\mathsf R(0)=R$.  For $u\ne0$, $\mathsf R(u)$ is a counting-field
deformation rather than a physical transition matrix.

The essential property is
\begin{equation}
 \mathsf R(u)
 \ne
 R\operatorname{diag}\!\left[\chi_+(u),1\right]
 \label{eq:no-factorization}
\end{equation}
in general.  Thus the work generated during the driven branch and the next
measurement outcome are not conditionally independent.  Equation~
\eqref{eq:no-factorization} is the mechanism by which the local
coherence--transition sector is transferred to the correlated outcome
record.

\begin{figure}[h]
\vspace*{0.3cm}  % 在这里增加垂直间距，数值可调，例如 0.8cm 或 1cm
\vspace{0.15cm}
\hspace*{0.1cm}
\begin{overpic}[width=3.3cm]
{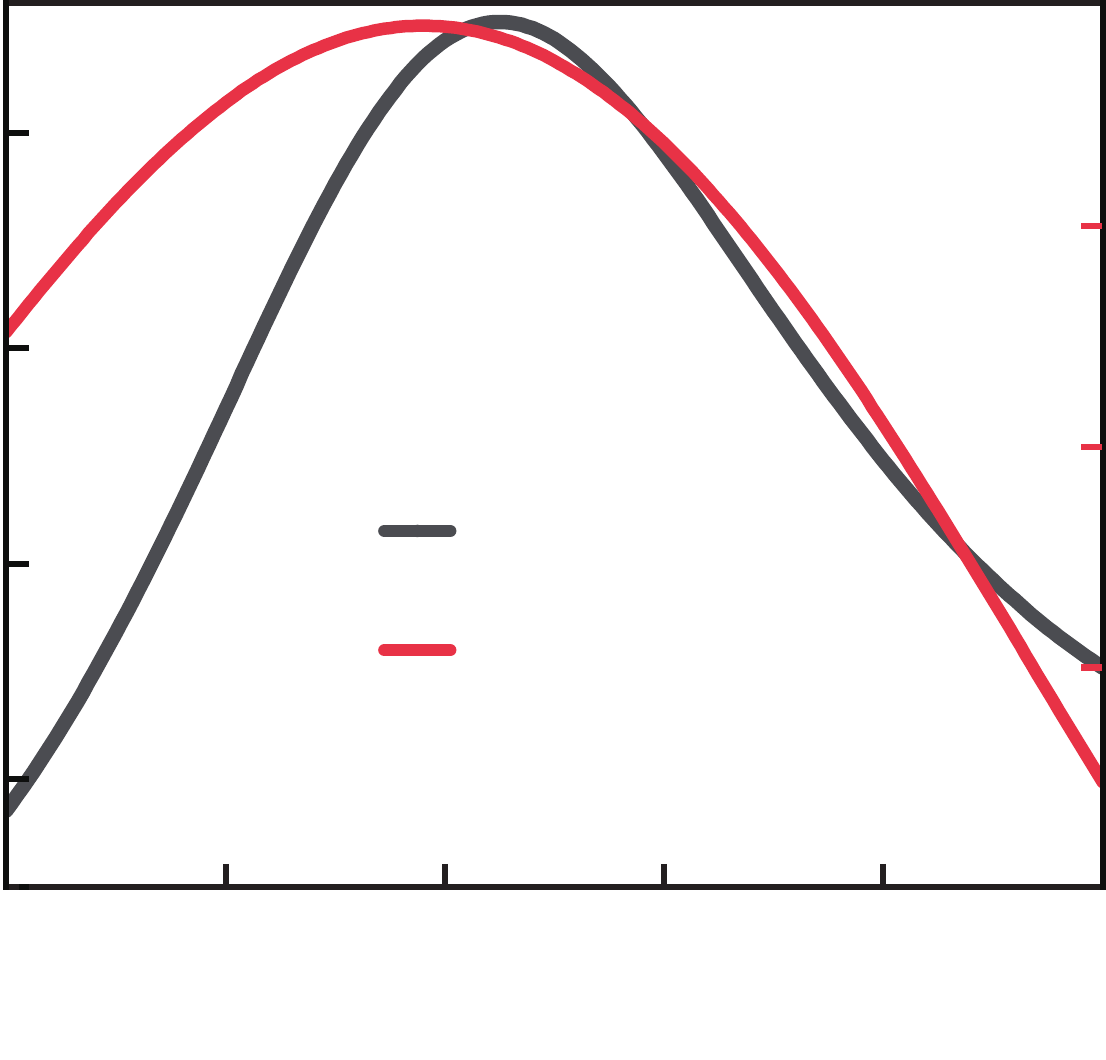}
\put(0,100){\scalebox{1}{(a)}}
\put(44,-6){\scalebox{1}{$\theta/\pi$}}
\put(45,47){\scalebox{0.8}{$p_+^*$}}
\put(45,35){\scalebox{0.8}{$\lambda_2$}}
\put(-13,93){\scalebox{0.7}{0.75}}
\put(-10,82.5){\scalebox{0.7}{0.7}}
\put(-10,62){\scalebox{0.7}{0.6}}
\put(-10,43){\scalebox{0.7}{0.5}}
\put(-10,23){\scalebox{0.7}{0.4}}
\put(0,7){\scalebox{0.7}{0}}
\put(15,7){\scalebox{0.7}{0.1}}
\put(35,7){\scalebox{0.7}{0.2}}
\put(55,7){\scalebox{0.7}{0.3}}
\put(75,7){\scalebox{0.7}{0.4}}
\put(95,7){\scalebox{0.7}{0.5}}
\put(102,93){\scalebox{0.7}{0.6}}
\put(102,73){\scalebox{0.7}{0.4}}
\put(102,53){\scalebox{0.7}{0.2}}
\put(102,33){\scalebox{0.7}{0.0}}
\put(102,14){\scalebox{0.7}{-0.2}}
\end{overpic}
\centering
\hspace{1.2cm} 
\vspace{0.15cm}
\begin{overpic}[width=3.3cm]
{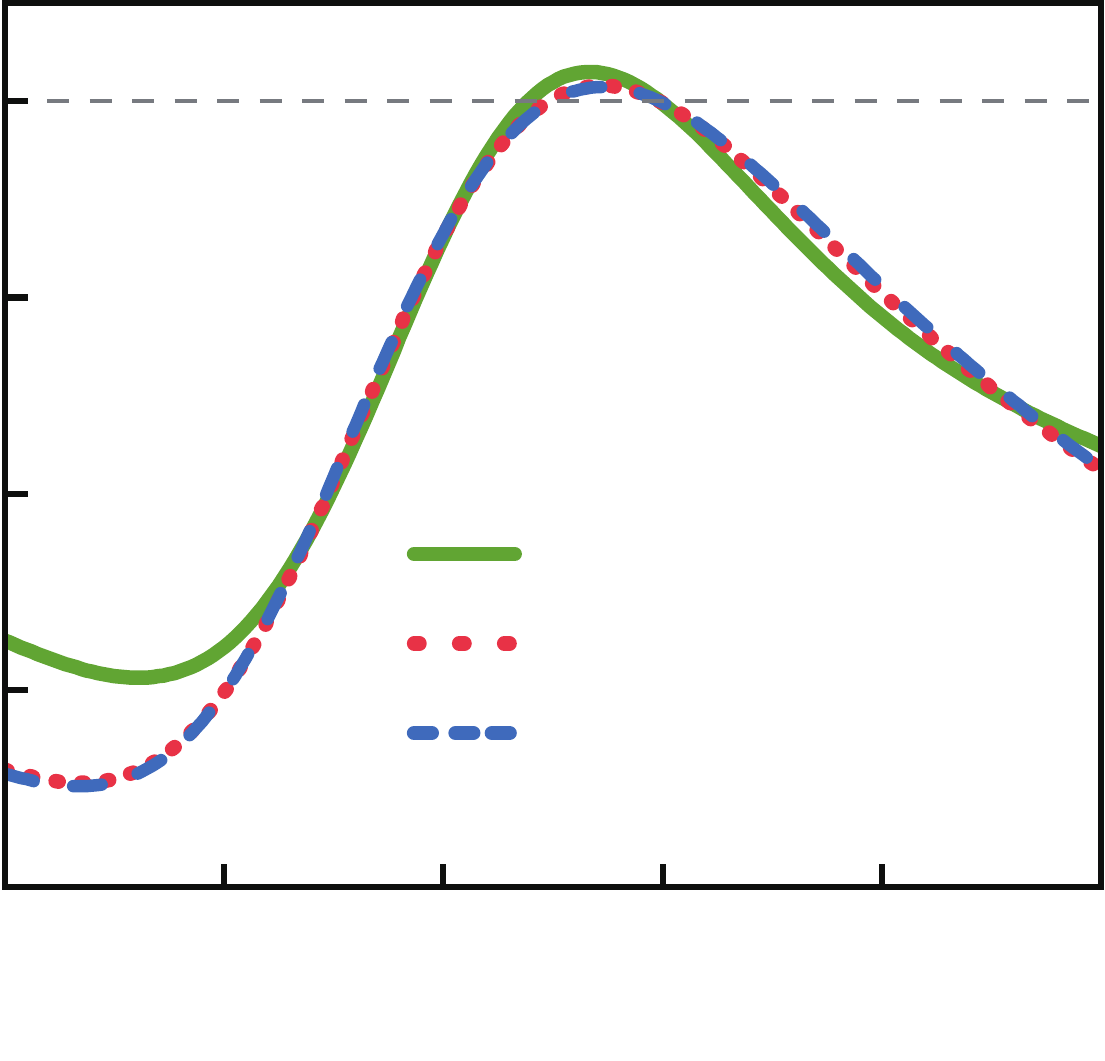}
\put(0,100){\scalebox{1}{(b)}}
\put(-22,40){\rotatebox{90}{\scalebox{0.8}{$\sqrt{\kappa_{2,\infty}/\kappa_{2,1}}$}}}
\put(44,-6){\scalebox{1}{$\theta/\pi$}}
\put(50,45){\scalebox{0.7}{$\tau_d=0.5$}}
\put(50,37){\scalebox{0.7}{$\tau_d=2$}}
\put(50,28){\scalebox{0.7}{$\tau_d=6$}}
\put(0,7){\scalebox{0.7}{0}}
\put(15,7){\scalebox{0.7}{0.1}}
\put(35,7){\scalebox{0.7}{0.2}}
\put(55,7){\scalebox{0.7}{0.3}}
\put(75,7){\scalebox{0.7}{0.4}}
\put(95,7){\scalebox{0.7}{0.5}}
\put(-10,14){\scalebox{0.7}{0.6}}
\put(-10,31.5){\scalebox{0.7}{0.7}}
\put(-10,49){\scalebox{0.7}{0.8}}
\put(-10,67){\scalebox{0.7}{0.9}}
\put(-7,85){\scalebox{0.7}{1}}
\put(-13,93){\scalebox{0.7}{1.05}}
\end{overpic}
\addtolength{\abovecaptionskip}{-3pt}
\caption{Outcome memory and accumulated-work fluctuations.
 (a) Stationary driven-branch probability $p_+^*$ and subleading memory
 eigenvalue $\lambda_2$ for $\tau_d=6$.
 (b) Ratio $\sqrt{\kappa_{2,\infty}/\kappa_{2,1}}$ for
 $\tau_d=0.5$, $2$, and $6$. {
The horizontal dashed line denotes the
independent-work-increment baseline
$\kappa_{2,\infty}/\kappa_{2,1}=1$.}  The full tilted kernel retains both the
 conditional FCS fluctuations and their correlations with subsequent
 outcomes.}
\label{fig:memory}
\vspace{0cm}
\end{figure}
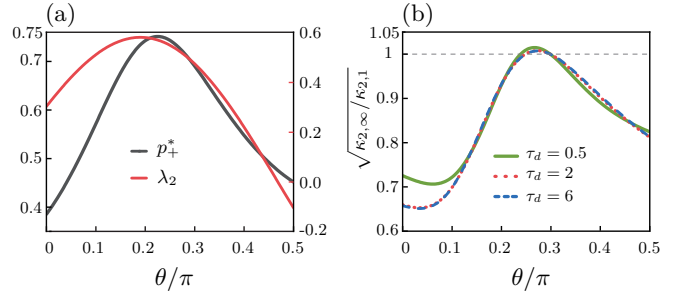

For \(N\) completed cycles, let \(\bm p^{(0)}\) denote the distribution of the initial outcome \(a_0\), which selects the first counted feedback branch, and let \(\bm 1=(1,1)^{\mathsf T}\) denote the all-ones vector, so that \(\bm 1^{\mathsf T}\) sums over the final outcome \(a_N\). The corresponding FCS generating function is
\begin{equation}
 \mathcal G_N(u)
 =\one^{\mathsf T}\mathsf R(u)^N\bm p^{(0)}.
 \label{eq:finite-N-generating}
\end{equation}
For stationary initialization, \(\bm p^{(0)}=\bm p^*\).
For the ergodic binary kernel, let $\Lambda_+(u)$ and $\Lambda_-(u)$ denote
the two eigenvalue branches analytic near $u=0$, with
$\Lambda_+(0)=1$ and $\Lambda_-(0)=\lambda_2$.  In this neighborhood, the
exact finite-cycle result has the two-mode form
\begin{equation}
 \mathcal G_N(u)
 =A_+(u)\Lambda_+(u)^N+A_-(u)\Lambda_-(u)^N.
 \label{eq:two-mode-generating}
\end{equation}

The dominant branch generates the extensive long-cycle statistics, while the
subleading branch carries the decaying outcome-memory mode.  With
$\psi(u)=\ln\Lambda_+(u)$, the cumulants per completed cycle are
\cite{Esposito2009}
\begin{equation}
 \kappa_{m,\infty}
 =(-\ii)^m\left.\partial_u^m\psi(u)\right|_{u=0}.
 \label{eq:scaled-cumulants}
\end{equation}
The corresponding finite-$N$ cumulants separate as
\begin{align}
 \kappa_m^{(N)}
 &=N\kappa_{m,\infty}+\mathcal B_m+\delta_{m,N},\nonumber\\
 \delta_{m,N}&=\mathcal O\!\left(N^{m-1}\varrho^N\right),
 \qquad |\lambda_2|<\varrho<1.
 \label{eq:finite-N-asymptotic}
\end{align}
Here $\mathcal B_m$ is an $N$-independent boundary contribution, whereas
$\delta_{m,N}$ is the exponentially decaying memory contribution.  For
stationary initialization, the mean is exactly extensive for every $N$;
higher cumulants generally retain the boundary term.

{
For $m=2$, the stationary binary outcome process
yields the exact finite-$N$ second cumulant
}
{
\begin{align}
\kappa_2^{(N)}
&=N\kappa_{2,\infty}
-\frac{2A(1-\lambda_2^N)}{(1-\lambda_2)^2},
\nonumber\\
\kappa_{2,\infty}
&=\kappa_{2,1}+\frac{2A}{1-\lambda_2},
\label{eq:exact-binary-cumulant}
\end{align}
}{
where
$A=\mathbf{1}^{\mathsf T}\mathsf M_1(\id-\mathcal{P}_*)\mathsf M_1\bm p^*$,
$\mathcal{P}_*=\bm p^*\mathbf{1}^{\mathsf T}$, and
$\mathsf M_1=\left.\partial_k \mathsf R(-ik)\right|_{k=0}$.
 The coefficient $A$ quantifies the connected FCS
work correlation between adjacent cycles.
In particular, $\lambda_2=0$ gives
$\kappa_{2,\infty}=\kappa_{2,1}+2A$,
which need not equal $\kappa_{2,1}$.
Thus independent measurement outcomes do not
necessarily imply independent work increments.
The correction can either suppress or enhance
the accumulated-work fluctuations.
}The energy-basis
construction of Eq.~\eqref{eq:resolved-instrument} and the detailed
derivations of Eqs.~\eqref{eq:finite-N-generating},
\eqref{eq:two-mode-generating}, ~\eqref{eq:finite-N-asymptotic}, ~\eqref{eq:exact-binary-cumulant} are given
in Appendix~\ref{app:cycles}.

As shown in Fig.~\ref{fig:memory}(a), \(p_+^*\) and \(\lambda_2\)
characterize distinct properties of the stationary engine.  The probability
\(p_+^*\) is the stationary fraction of cycles following the driven branch
and therefore weights the single-cycle work statistics.  By contrast,
\(\lambda_2\) governs the outcome correlations through
\(C_I(\ell)/C_I(0)=\lambda_2^\ell\): its magnitude sets their decay, while
its sign distinguishes persistent from alternating records.  Thus the
stationary branch composition does not determine the temporal memory.

The resulting work-fluctuation correction is quantified in
Fig.~\ref{fig:memory}(b) by
\(\sqrt{\kappa_{2,\infty}/\kappa_{2,1}}\).  The denominator has the same stationary one-cycle
marginal but excludes intercycle correlations,
so values below (above) unity indicate
{
suppression (enhancement) due to intercycle
work correlations.
}
Suppression occurs over most of the plotted range,
with a narrow intermediate region of enhancement.
{
Such correlations can persist even at $\lambda_2=0$
through work--next-outcome correlations, as shown
by Eq.~\eqref{eq:exact-binary-cumulant}.
}

\begin{figure}[h]
\vspace*{0.2cm}  % 在这里增加垂直间距，数值可调，例如 0.8cm 或 1cm
\vspace{0cm}
\hspace*{0.5cm}
\begin{overpic}[width=3.3cm]
{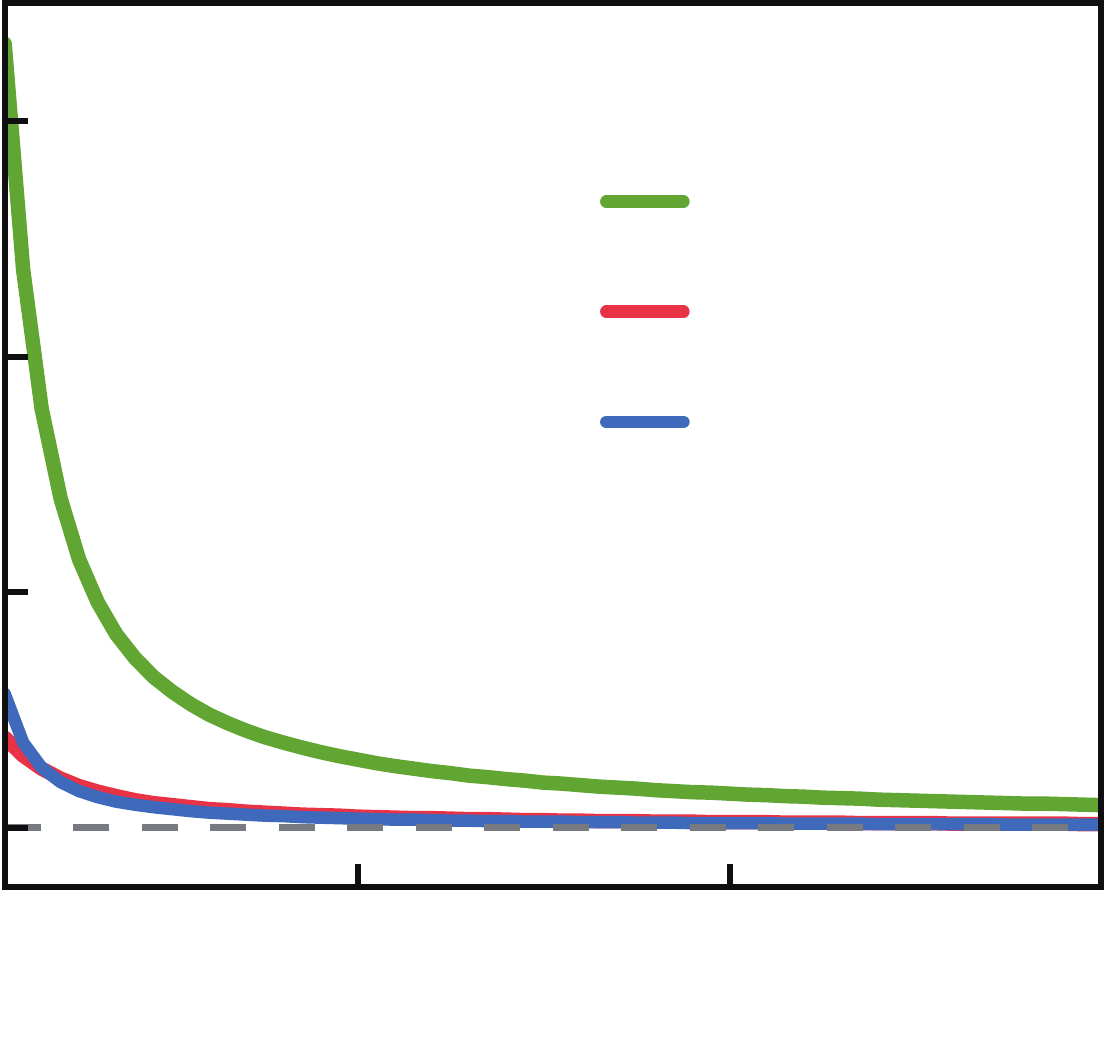}
\put(0,100){\scalebox{0.9}{(a)}}
\put(-24,36){\rotatebox{90}{\scalebox{0.7}{$\kappa_2^{(N)}/(N\kappa_{2,\infty})$}}}
\put(44,-3){\scalebox{1}{$N$}}
\put(65,76){\scalebox{0.7}{$\theta=0.05\pi$}}
\put(65,66){\scalebox{0.7}{$\theta=0.2\pi$}}
\put(65,56){\scalebox{0.7}{$\theta=0.4\pi$}}
\put(-10,93){\scalebox{0.7}{2.4}}
\put(-10,82.5){\scalebox{0.7}{2.2}}
\put(-10,62){\scalebox{0.7}{1.8}}
\put(-10,41){\scalebox{0.7}{1.4}}
\put(-7,19.5){\scalebox{0.7}{1}}
\put(-1,7){\scalebox{0.7}{1}}
\put(29.5,7){\scalebox{0.7}{20}}
\put(63.5,7){\scalebox{0.7}{40}}
\put(95,7){\scalebox{0.7}{60}}
\end{overpic}
\centering
\hspace{0.9cm} 
\vspace{0.15cm}
\begin{overpic}[width=3.3cm]
{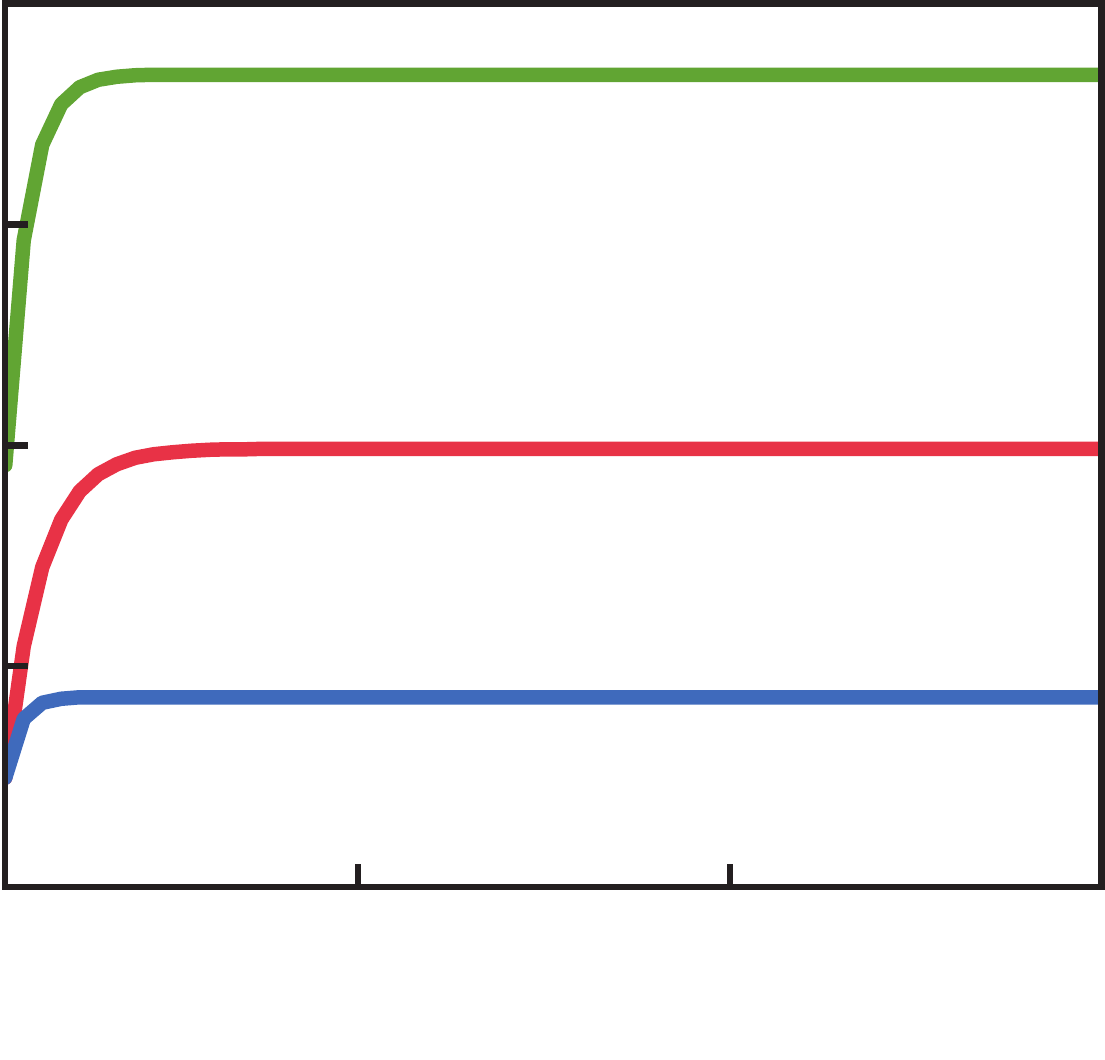}
\put(0,100){\scalebox{0.9}{(b)}}
\put(44,-3){\scalebox{1}{$N$}}
\put(-26,30){\rotatebox{90}{\scalebox{0.7}{$\kappa_2^{(N)}-N\kappa_{2,\infty}$}}}
\put(-13,93){\scalebox{0.7}{0.18}}
\put(-13,74){\scalebox{0.7}{0.14}}
\put(-10,53.5){\scalebox{0.7}{0.1}}
\put(-13,34){\scalebox{0.7}{0.06}}
\put(-13,15){\scalebox{0.7}{0.02}}
\put(-1,7){\scalebox{0.7}{1}}
\put(29.5,7){\scalebox{0.7}{20}}
\put(63.5,7){\scalebox{0.7}{40}}
\put(95,7){\scalebox{0.7}{60}}
\end{overpic}
\addtolength{\abovecaptionskip}{-7pt}
\caption{Exact finite-cycle convergence at $\tau_d=2$ for stationary
 initialization.  (a) Per-cycle second FCS cumulant normalized by its
 long-cycle value.  (b) Boundary difference
 $\kappa_2^{(N)}-N\kappa_{2,\infty}$.  The plateau is the boundary
 contribution; the remaining memory contribution decays exponentially.
 Work is measured in units of $\epsc$.}
\label{fig:finite-cycle}
\vspace{0cm}
\end{figure}
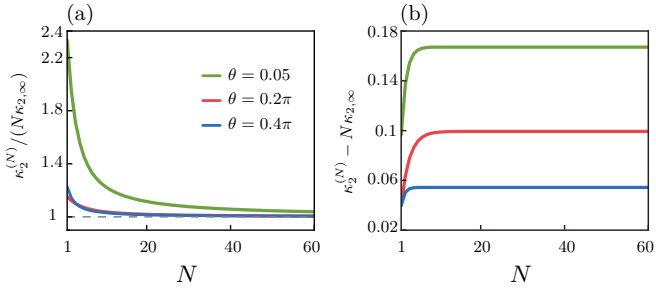

The approach to the correlated long-cycle limit is resolved in
Fig.~\ref{fig:finite-cycle}.  In
Fig.~\ref{fig:finite-cycle}(a),
\(\kappa_2^{(N)}/(N\kappa_{2,\infty})\) approaches unity as the number of
completed cycles increases.  Figure~\ref{fig:finite-cycle}(b) isolates the
finite-cycle correction: the exponentially decaying memory contribution
vanishes with \(N\), whereas the boundary contribution approaches the
nonzero plateau \(\mathcal B_2\).  Thus outcome memory controls the transient
approach to the long-cycle statistics, {while the boundary term
can remain nonzero even for stationary initialization.}

\subsection{Work fluctuations at fixed laboratory time}
\label{sec:physical-time}

The preceding statistics refer to a fixed number of completed cycles \cite{Menczel2020}.  At a
fixed laboratory time, the number of completed cycles also fluctuates because
the outcome selects the branch duration,
\begin{equation}
 \tau_+=\tau_c+\tau_e,
 \qquad
 \tau_-=\tau_d.
 \label{eq:branch-durations}
\end{equation}
A Laplace field $\zeta$ for elapsed time gives the joint work--duration
kernel
\begin{align}
 \mathsf R(u,\zeta)&=\mathsf R(u)\mathsf D(\zeta),
 \nonumber\\
 \mathsf D(\zeta)
 &=\operatorname{diag}\!\left(
 \ee^{-\zeta\tau_+},\ee^{-\zeta\tau_-}
 \right),
 \label{eq:joint-kernel}
\end{align}
where the time factor multiplies the column associated with the branch that
has just been completed.

Let $\Lambda_0(u,\zeta)$ be the eigenvalue analytic near unity and define
$\Psi(u,\zeta)=\ln\Lambda_0(u,\zeta)$.  
Its derivatives at the origin give
the long-cycle joint cumulants
\begin{align}
 \bar w&=-\ii\Psi_u,
 &
 \bar\tau&=-\Psi_\zeta,
 \nonumber\\
 C_{ww}&=-\Psi_{uu},
 &
 C_{w\tau}&=\ii\Psi_{u\zeta},
 &
 C_{\tau\tau}&=\Psi_{\zeta\zeta}.
 \label{eq:joint-cumulants}
\end{align}
{Here, 
$\Psi_{uu}
\equiv
\left.
\frac{\partial^2\Psi}{\partial u^2}
\right|_{(u,\zeta)=(0,0)}$
and
$\Psi_{u\zeta}
\equiv
\left.
\frac{\partial^2\Psi}{\partial u\,\partial\zeta}
\right|_{(u,\zeta)=(0,0)}$. Analogously, $\Psi_u$, $\Psi_\zeta$, and
$\Psi_{\zeta\zeta}$ denote the corresponding first and
second partial derivatives at the origin.}
Here $\bar\tau=p_+^*\tau_++p_-^*\tau_-$,
$C_{ww}=\kappa_{2,\infty}$, $C_{w\tau}$ is the work--duration cross
cumulant, and $C_{\tau\tau}$ is the duration cumulant.  These quantities
already include the correlations generated by the outcome record.

Let \(W(t)\) denote the cumulative driven work accumulated over all feedback branches completed by laboratory time \(t\), with the cold-contact branch contributing zero to the counted-work ledger. In the long-time limit,
\(
\ln \left\langle e^{iuW(t)}\right\rangle_{\rm FC}
=t\phi(u)+O(1), 
\)
where the local cumulant-generating function \(\phi(u)\) is determined implicitly by

\begin{equation}
 \Lambda_0\!\left(u,\phi(u)\right)=1,
 \qquad
 \phi(0)=0.
 \label{eq:physical-time-root}
\end{equation}
Implicit differentiation yields the mean-work rate
\begin{equation}
 j_w\equiv-\ii\phi'(0)=\frac{\bar w}{\bar\tau}
 \label{eq:mean-work-rate}
\end{equation}
and the fixed-time second-cumulant rate
\begin{equation}
 D_w\equiv-\phi''(0)
 =\frac{C_{ww}-2j_wC_{w\tau}+j_w^2C_{\tau\tau}}
 {\bar\tau}.
 \label{eq:physical-time-second-cumulant}
\end{equation}
With the present sign convention, $-j_w$ is the extracted-work rate.  The
second-cumulant rate contains three physically distinct contributions: the
completed-cycle work fluctuation, a signed work--duration correction, and the
duration fluctuation.  The detailed derivations of
Eqs.~\eqref{eq:physical-time-root}, \eqref{eq:mean-work-rate},
and~\eqref{eq:physical-time-second-cumulant} are given in
Appendix~\ref{app:physical-time}.

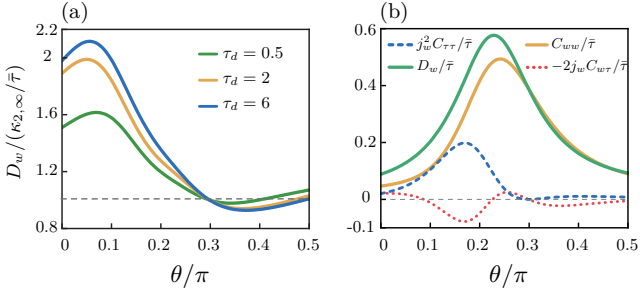
\begin{figure}[h]
\vspace*{0.1cm}  % 在这里增加垂直间距，数值可调，例如 0.8cm 或 1cm
\vspace{0.15cm}
\hspace*{0.5cm}
\begin{overpic}[width=3.3cm]
{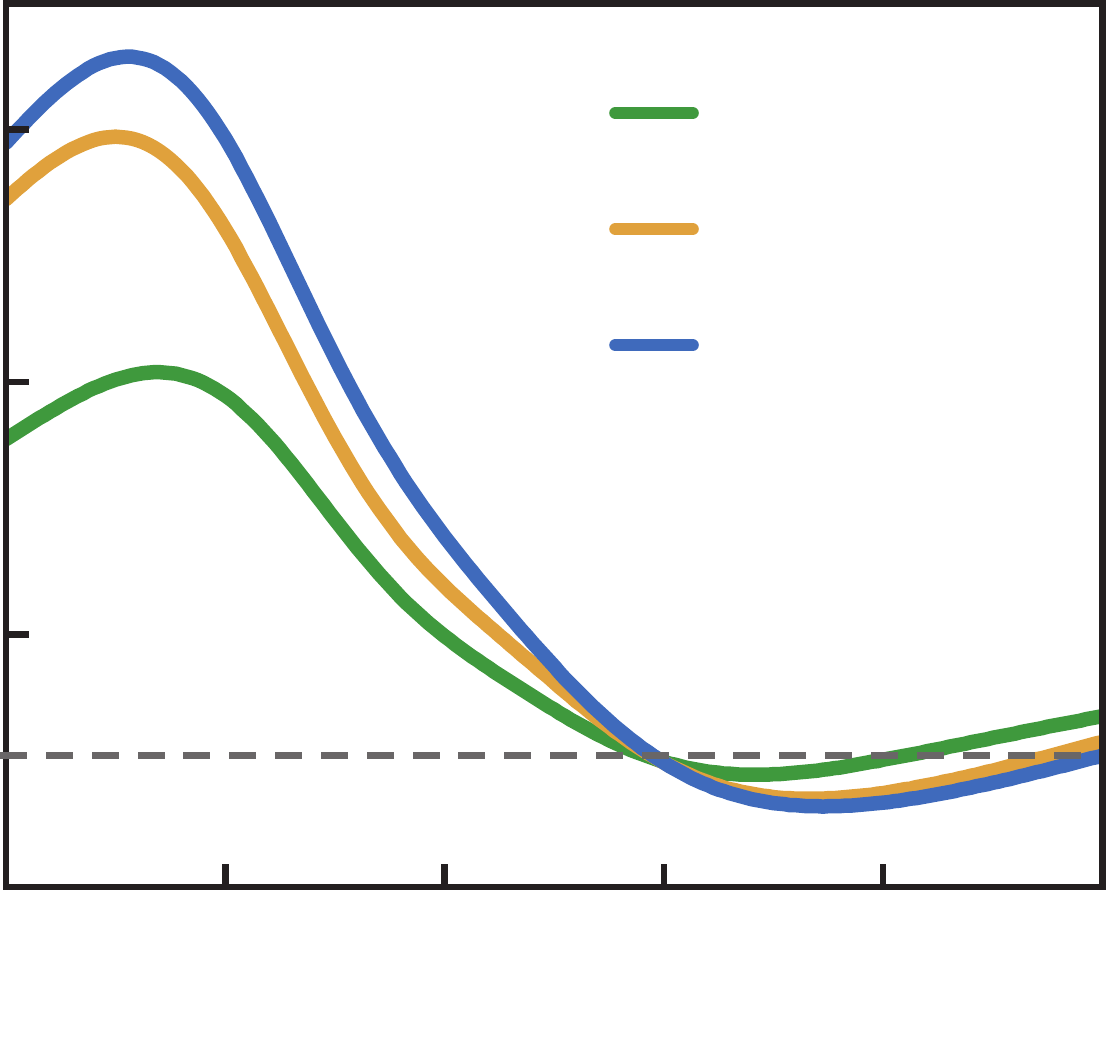}
\put(0,100){\scalebox{0.9}{(a)}}
\put(-22,35){\rotatebox{90}{\scalebox{0.8}{$D_w/(\kappa_{2,\infty}/\bar\tau)$}}}
\put(44,-6){\scalebox{1}{$\theta/\pi$}}
\put(65,84){\scalebox{0.7}{$\tau_d=0.5$}}
\put(65,74){\scalebox{0.7}{$\tau_d=2$}}
\put(65,64){\scalebox{0.7}{$\tau_d=6$}}
\put(-10,93){\scalebox{0.7}{2.2}}
\put(-7,82.5){\scalebox{0.7}{2}}
\put(-10,60){\scalebox{0.7}{1.6}}
\put(-10,37){\scalebox{0.7}{1.2}}
\put(-7,26){\scalebox{0.7}{1}}
\put(-10,14){\scalebox{0.7}{0.8}}
\put(0,7){\scalebox{0.7}{0}}
\put(15,7){\scalebox{0.7}{0.1}}
\put(35,7){\scalebox{0.7}{0.2}}
\put(55,7){\scalebox{0.7}{0.3}}
\put(75,7){\scalebox{0.7}{0.4}}
\put(95,7){\scalebox{0.7}{0.5}}
\end{overpic}
\centering
\hspace{0.7cm} 
\vspace{0.15cm}
\begin{overpic}[width=3.3cm]
{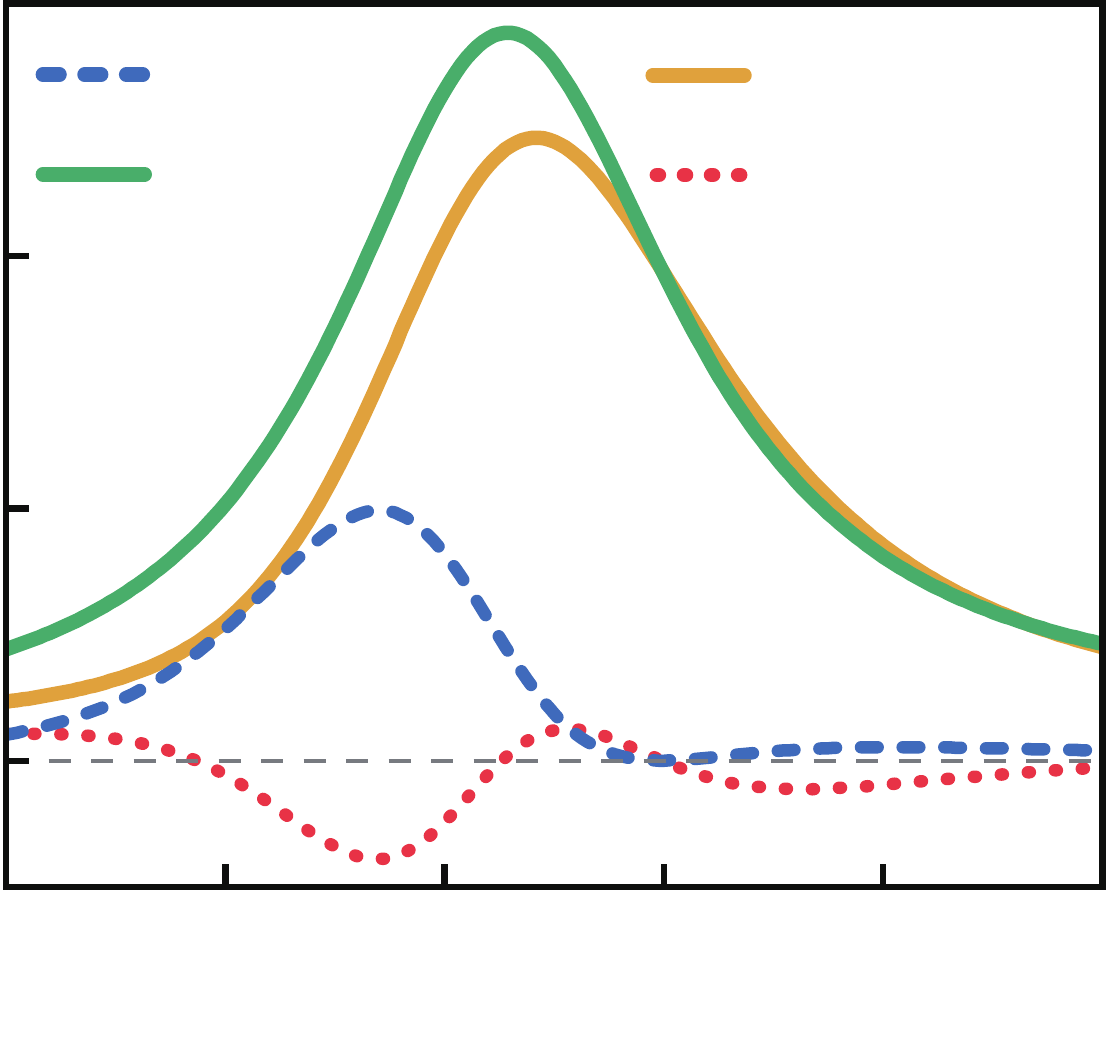}
\put(0,100){\scalebox{0.9}{(b)}}
\put(44,-6){\scalebox{1}{$\theta/\pi$}}
\put(69,88){\scalebox{0.6}{$C_{ww}/\bar\tau$}}
\put(69,79){\scalebox{0.6}{$-2j_wC_{w\tau}/\bar\tau$}}
\put(15,88){\scalebox{0.6}{$j_w^2C_{\tau\tau}/\bar\tau$}}
\put(15,79){\scalebox{0.6}{$D_w$}}
\put(0,7){\scalebox{0.7}{0}}
\put(15,7){\scalebox{0.7}{0.1}}
\put(35,7){\scalebox{0.7}{0.2}}
\put(55,7){\scalebox{0.7}{0.3}}
\put(75,7){\scalebox{0.7}{0.4}}
\put(95,7){\scalebox{0.7}{0.5}}
\put(-13,14){\scalebox{0.7}{-0.1}}
\put(-7,26){\scalebox{0.7}{0}}
\put(-10,48){\scalebox{0.7}{0.2}}
\put(-10,71){\scalebox{0.7}{0.4}}
\put(-10,93){\scalebox{0.7}{0.6}}
\end{overpic}
\caption{Fixed-time symmetric-FCS second-cumulant rate.
 (a) Exact $D_w$ divided by the naive completed-cycle rescaling
 $\kappa_{2,\infty}/\bar\tau$ for $\tau_d=0.5$, $2$, and $6$.
 The horizontal dashed line denotes equality.
 (b) Completed-cycle, work--duration, and duration contributions in
 Eq.~\eqref{eq:physical-time-second-cumulant} at $\tau_d=2$.
 Work and time are measured in units of $\epsc$ and $\gamma^{-1}$,
 respectively.}
 \label{fig:physical-time}
\vspace{0cm}
\end{figure}

If \(\tau_+=\tau_-\), fixed cycle number also fixes the elapsed time, so
\(C_{w\tau}=C_{\tau\tau}=0\) and
\(D_w=C_{ww}/\bar\tau\).  When the branch durations differ, the number of
cycles completed within a fixed time fluctuates and the work--duration and
duration terms must be retained.  As seen in
Fig.~\ref{fig:physical-time}(a), the exact rate can lie above or below the
naive rescaling \(\kappa_{2,\infty}/\bar\tau\), corresponding respectively
to its underestimation or overestimation.

The decomposition in Fig.~\ref{fig:physical-time}(b) identifies the origin of
this difference: the nonnegative duration term and the signed work--duration
term can reinforce or partially cancel each other.  This is a distinction
between work fluctuations sampled at fixed cycle number and at fixed
laboratory time, not an additional change in mean power or efficiency.

\section{Record cost and operational scope}
\label{sec:reset}

{
Measurement engines and memory-reset processes
have been studied from complementary thermodynamic
perspectives \cite{Jordan2020,ReebWolf2014,Liu2025Reset}.
The outcome transition matrix $R$ underlying
the tilted work kernel $\mathsf R(u)$ also determines
the measurement-record entropy rate and its minimum
Landauer reset cost.
Distinct from the working-medium heat exchange,
this cost concerns the controller memory.
For degenerate logical memory states, reversible
isothermal erasure requires
$Q_{\mathrm{reset}}^{\min}=k_B T S_{\mathrm{mem}}$,
equal to the heat released to the reset reservoir.
Here, $S_{\mathrm{mem}}$ is the dimensionless memory
entropy, and $T$ is the reset-reservoir temperature,
which need not equal the cold-bath temperature $T_c$.
}
For a stationary first-order Markov record, define
\begin{align}
 h_\mu&=-\sum_{a,b}p_a^*q_{b|a}\ln q_{b|a},
 \label{eq:entropy-rate}\\
 H_2(p)&=-p\ln p-(1-p)\ln(1-p).
\end{align}
{
Here, $h_\mu=H(X_{n+1}|X_n)$ is the conditional
Shannon entropy and hence the entropy rate
of the stationary measurement record, measured
in nats per cycle. }

{
For a length-$N$ outcome record
$X^N=(X_1,\ldots,X_N)$,
$Q_X^{(N)}=k_B T H(X^N)$ denotes the minimum
total reset cost, with $Q_X^{(N)}/N$ giving
the corresponding per-cycle cost.
For the stationary first-order Markov record, we obtain
}
\begin{align}
 H(X_1,\ldots,X_N)&=H_2(p_+^*)+(N-1)h_\mu,
 \label{eq:block-entropy}\\
 \frac{Q_X^{(N)}}{N}
 &=\kb T\left[h_\mu+\frac{H_2(p_+^*)-h_\mu}{N}\right].
 \label{eq:block-cost}
\end{align}

{
Equation~\eqref{eq:block-cost}
shows that buffering reduces the per-cycle cost
from $Q_X^{(1)}=k_B T H_2(p_+^*)$ to
$Q_X^{(\infty)}=k_B T h_\mu$, where $Q_X^{(\infty)}$ is the asymptotic
per-cycle erasure cost of the outcome record.
The finite-$N$ correction is given exactly
by Eq.~\eqref{eq:block-cost}.
This reduction arises from temporal correlations
in the outcome record, rather than changes in
the working-medium heat exchange.}
The corresponding predictive
information is \cite{Still2012}
\begin{equation}
 H_2(p_+^*)-h_\mu=I(X_n:X_{n-1}).
 \label{eq:excess-entropy}
\end{equation}
{At $\lambda_2=0$, $h_\mu=H_2(p_+^*)$, so the reset-cost
saving from buffering vanishes, although adjacent work increments can
remain correlated through $A\ne0$ in Eq.~\eqref{eq:exact-binary-cumulant}.}
These statements require a stationary source, a block buffer, reversible
compression and erasure, and degenerate logical memory states.  If a
separate state-bearing register is also archived, its conditional entropy
must be included.  The asymptotic joint cost is
\begin{equation}
 Q_{XS}^{(\infty)}
 =\kb T\left[h_\mu+\sum_a p_a^*S(\rho_a)\right],
 \label{eq:joint-cost}
\end{equation}
where $S(\rho)=-\Tr(\rho\ln\rho)$. 
{
The quantity $Q_{XS}^{(\infty)}$ denotes the
asymptotic per-cycle reset cost when both the outcome
record $X$ and a separate state register $S$ are erased.
Its two terms represent the entropy-rate cost of $X$
and the conditional reset cost of $S$, respectively.
Without access to $X$, resetting $S$ instead costs
$Q_S=k_B T S(\bar{\rho})$, where
$\bar{\rho}=\sum_a p_a^*\rho_a$.
These conditional and unconditional reset costs
correspond to alternative protocols.
}
 The detailed derivations of Eqs.~\eqref{eq:block-entropy},
\eqref{eq:block-cost}, \eqref{eq:excess-entropy}, and~\eqref{eq:joint-cost},
together with the distinction between the physical reset ledgers, are given
in Appendix~\ref{app:reset}.

The required quantities separate operationally.  Outcome counts determine
$ R$, $\bm p^*$, $\lambda_2$, and $h_\mu$ without an energy measurement.  By
contrast, the coherent work sector requires $\mathcal T_{\rm ce}$ and the
gauge-invariant phase $\phi_{\rm rel}$, obtainable from counting-field
interferometry \cite{Dorner2013,Mazzola2013}, a suitable weak-measurement scheme
\cite{Gherardini2024}, or calibrated state and process tomography.
Two drives with the same $\mathcal T_{\rm ce}$ can therefore have different
coherent corrections.  Thus the record statistics and the local coherent
work sector require complementary measurements.

\section{Conclusions}
\label{sec:conclusions}

Measurement-induced coherence enters the finite-time work statistics through its
phase-sensitive interference with the transitions generated by the driven branch.
The measurement prepares a population bias and energy coherence, while the
complete driven propagator supplies the endpoint mixing $\mathcal T_{ce}$.
This interference is quantified by $\nu_{\rm FC}$.  For the symmetric FCS ordering,
$\mathcal T_{\rm ce}$ fixes every even raw work moment of the unitary qubit
branch, whereas $\nu_{\rm FC}$ enters the odd moments through an
equal-and-opposite half-quantum pair.  The second raw moment therefore has no
independent coherent term, although the second cumulant remains coherence
dependent through the mean.  Coherence can enhance or suppress both work
extraction and fluctuations; its magnitude alone is insufficient.

The energetic meaning of endpoint mixing depends on the input state: it is a
nonnegative inner-friction cost for the passive Gibbs reference, but it can
release population-inversion energy from the active measurement-prepared
state, with the coherent part reinforcing or opposing that release.

During repeated operation, the local interference is sampled by a correlated
outcome record.  The outcome-resolved tilted kernel retains both the
conditional FCS and its correlation with the following outcome.  It separates
finite-cycle corrections into an algebraic boundary term and a decaying
memory contribution.  Unequal branch durations further generate
work--duration and duration terms in the fixed-time second-cumulant rate, so a
simple division of the cycle-count cumulant by the mean duration is generally
incorrect.  {
Temporal correlations in the outcome record
also reduce the minimum per-cycle reset cost
of a buffered measurement record from
$k_B T H_2(p_+^*)$ to $k_B T h_\mu$.
This provides an information-thermodynamic
counterpart to the work-statistics results.
}  Together, these results identify an experimentally
accessible coherence signature in the work quasiprobability and establish
how it survives in correlated finite-cycle and fixed-time fluctuations.

\begin{acknowledgments}
This work was supported by the National Natural Science Foundation of China
(Grant No. 12465009).  J.W. additionally acknowledges support from the Major
Program of Jiangxi Provincial Natural Science Foundation, China (Grant No.
20224ACB201007).
\end{acknowledgments}

\begin{center}
\textbf{DATA AVAILABILITY}
\end{center}

There are no publicly available research data or software
supporting this manuscript. Requests for further information
or data should be sent to the authors.

\appendix

\section{Endpoint work identities and five-point full counting statistics}
\label{app:fcs}

The endpoint expansion below is used to derive
Eqs.~\eqref{eq:measurement-interference-general},
\eqref{eq:work-mechanism},
\eqref{eq:measurement-interference}, and
\eqref{eq:inner-friction-reference} of the main text.  The subsequent
energy-basis expansion of the symmetric FCS characteristic function is used
to derive Eqs.~\eqref{eq:five-weights},
\eqref{eq:negativity-identity}, and
\eqref{eq:moment-parity};
Eqs.~\eqref{eq:conditional-moments} and
\eqref{eq:coherence-cumulant-correction} then follow as direct consequences.

In the ordered energy basis $(|e\rangle,|g\rangle)$, an arbitrary initial
qubit state can be written as
\begin{equation}
 \rho=
 \begin{pmatrix}
  \rho_{ee}&\rho_{eg}\\
  \rho_{ge}&\rho_{gg}
 \end{pmatrix},
 \qquad
 \rho_{eg}=\rho_{ge}^*,
 \qquad
 \rho_{ee}+\rho_{gg}=1.
 \label{eq:app-fcs-initial-state}
\end{equation}
For brevity, $U\equiv U_{\rm ce}$ is used in this appendix.  The final
excited-state probability is
\begin{align}
 p_e^f(\rho)
 &=\langle e|U\rho U^\dagger|e\rangle\nonumber\\
 &=|U_{ee}|^2\rho_{ee}+|U_{eg}|^2\rho_{gg}
 +2\operatorname{Re}\!\left(U_{eg}U_{ee}^*\rho_{ge}\right).
 \label{eq:app-final-excited-population}
\end{align}
Complete dephasing in the $H_0$ eigenbasis removes only the last term.
Consequently,
\begin{equation}
 p_e^f(\rho)-p_e^f\!\left(\Delta_0[\rho]\right)
 =2\operatorname{Re}\!\left(U_{eg}U_{ee}^*\rho_{ge}\right),
 \label{eq:app-coherence-population-change}
\end{equation}
which proves Eq.~\eqref{eq:measurement-interference-general}.  No purity
condition has been used.

Because the initial and final Hamiltonians of the complete driven branch are
both $H_0$, its mean work is the endpoint energy change
\begin{align}
 \bar w[\rho]
 &=\Tr\!\left[H_0\left(U\rho U^\dagger-\rho\right)\right]\nonumber\\
 &=\epsc\left[p_e^f(\rho)-\rho_{ee}\right].
 \label{eq:app-endpoint-work}
\end{align}
Unitarity gives
$|U_{eg}|^2=|U_{ge}|^2=\mathcal T_{\rm ce}$ and
$|U_{ee}|^2=|U_{gg}|^2=1-\mathcal T_{\rm ce}$.  Inserting
Eq.~\eqref{eq:app-final-excited-population} into
Eq.~\eqref{eq:app-endpoint-work} therefore yields
\begin{equation}
 \bar w[\rho]
 =\epsc\left[
 \mathcal T_{\rm ce}(\rho_{gg}-\rho_{ee})
 +\nu_{\rm FC}[\rho]
 \right].
 \label{eq:app-general-mean-work}
\end{equation}
For $\rho=\Pi_+$,
$\rho_{ee}-\rho_{gg}=\cos\theta$ and
$\nu_{\rm FC}\equiv\nu_{\rm FC}[\Pi_+]$.  Equation
\eqref{eq:app-general-mean-work} then gives the population and coherent
contributions stated in Eq.~\eqref{eq:work-mechanism}.

When
$U_{eg}U_{ee}^*\rho_{ge}\ne0$, its modulus--phase decomposition gives
\begin{equation}
 \nu_{\rm FC}[\rho]
 =2|\rho_{ge}|
 \sqrt{\mathcal T_{\rm ce}(1-\mathcal T_{\rm ce})}
 \cos\phi_{\rm rel}.
 \label{eq:app-interference-phase-form}
\end{equation}
For $\Pi_+$ in the range considered in the main text,
$2|\rho_{ge}|=\sin\theta$, and
Eq.~\eqref{eq:measurement-interference} follows.  If the product vanishes,
$\nu_{\rm FC}=0$ directly and the phase need not be defined.

The energetic meaning of $\mathcal T_{\rm ce}$ can be checked with the
passive reference
$\rho_\beta=\ee^{-\beta H_0}/\Tr(\ee^{-\beta H_0})$.  Since this state is
energy diagonal,
$\nu_{\rm FC}[\rho_\beta]=0$, while
$(\rho_\beta)_{gg}-(\rho_\beta)_{ee}
=\tanh(\beta\epsc/2)$.  Equation~\eqref{eq:app-general-mean-work} gives
\begin{equation}
 \Tr\!\left[H_0\left(U\rho_\beta U^\dagger-\rho_\beta\right)\right]
 =\epsc\mathcal T_{\rm ce}\tanh(\beta\epsc/2).
 \label{eq:app-passive-endpoint-work}
\end{equation}
Moreover, unitary invariance of the von Neumann entropy and
$\ln\rho_\beta=-\beta H_0-\ln Z_\beta$ imply
\begin{align}
 D(U\rho_\beta U^\dagger\Vert\rho_\beta)
 &=\Tr\!\left[U\rho_\beta U^\dagger
 \left(\ln U\rho_\beta U^\dagger-\ln\rho_\beta\right)\right]
 \nonumber\\
 &=\beta\Tr\!\left[H_0
 \left(U\rho_\beta U^\dagger-\rho_\beta\right)\right].
 \label{eq:app-relative-entropy-work}
\end{align}
Equations~\eqref{eq:app-passive-endpoint-work} and
\eqref{eq:app-relative-entropy-work} prove
Eq.~\eqref{eq:inner-friction-reference}.

For general endpoint Hamiltonians
$H_i=\sum_nE_n^i|n\rangle\langle n|$ and
$H_f=\sum_\ell E_\ell^f|\ell\rangle\langle\ell|$, the symmetric FCS
characteristic function is
\begin{equation}
 \chi_{\rm FC}(u)=\Tr\!\left[
 \ee^{\ii uH_f/2}U\ee^{-\ii uH_i/2}\rho
 \ee^{-\ii uH_i/2}U^\dagger\ee^{\ii uH_f/2}
 \right].
 \label{eq:app-general-fcs}
\end{equation}
Insertion of the initial and final energy resolutions gives
\begin{equation}
 \chi_{\rm FC}(u)=
 \sum_{\ell,n,n'}
 \ee^{\ii u[E_\ell^f-(E_n^i+E_{n'}^i)/2]}
 U_{\ell n}\rho_{nn'}U_{\ell n'}^*.
 \label{eq:app-general-fcs-expansion}
\end{equation}
Thus the corresponding work quasiprobability is
\begin{multline}
 \mathcal P_{\rm FC}(w)=
 \sum_{\ell,n,n'}U_{\ell n}\rho_{nn'}U_{\ell n'}^*\\
 \times
 \delta\!\left(
 w-E_\ell^f+\frac{E_n^i+E_{n'}^i}{2}
 \right).
 \label{eq:app-general-fcs-quasiprobability}
\end{multline}
The terms with $n=n'$ are the population-transfer sector.  The terms with
$n\ne n'$ contain the initial energy coherence and are assigned the midpoint
initial energy $(E_n^i+E_{n'}^i)/2$; their combined weights are real but need
not be positive.  Normalization follows from unitarity and $\Tr\rho=1$.

For the engine, $H_i=H_f=H_0$ and $U=U_{\rm ce}$.  The diagonal terms of
Eq.~\eqref{eq:app-general-fcs-quasiprobability} give work values
$+\epsc$, $-\epsc$, and $0$.  The off-diagonal pair has zero midpoint
initial energy and gives $+\epsc/2$ or $-\epsc/2$ according to the final
energy.  The five weights for an arbitrary qubit state are therefore
\begin{align}
 \mathcal P_{\rm FC}(+\epsc|{\rho})&=\mathcal T_{\rm ce}\rho_{gg},
 &
 \mathcal P_{\rm FC}(-\epsc|{\rho})&=\mathcal T_{\rm ce}\rho_{ee},
 \nonumber\\
 \mathcal P_{\rm FC}(0|{\rho})&=1-\mathcal T_{\rm ce},
 &
 \mathcal P_{\rm FC}(+\epsc/2|{\rho})&=\nu_{\rm FC}[\rho],
 \nonumber\\
 &&
 \mathcal P_{\rm FC}(-\epsc/2|{\rho})&=-\nu_{\rm FC}[\rho].
 \label{eq:app-general-five-weights}
\end{align}
For the last equality, column orthogonality has been used in the form
\begin{equation}
 U_{gg}U_{ge}^*=-U_{eg}U_{ee}^*.
 \label{eq:app-unitary-orthogonality}
\end{equation}
Substitution of
$\rho_{ee}=(1+\cos\theta)/2$ and
$\rho_{gg}=(1-\cos\theta)/2$ into
Eq.~\eqref{eq:app-general-five-weights} proves
Eq.~\eqref{eq:five-weights}.

The three integer-sector weights in
Eq.~\eqref{eq:app-general-five-weights} are nonnegative and sum to unity,
whereas the half-quantum pair has zero net weight.  Hence
\begin{align}
 \frac12\left[
 \sum_j|\mathcal P_{\rm FC}(w_j)|-1
 \right]
 &=|\nu_{\rm FC}|,\nonumber\\
 \sum_{|w_j|=\epsc/2}|\mathcal P_{\rm FC}(w_j)|
 &=2|\nu_{\rm FC}|.
 \label{eq:app-negativity-evaluation}
\end{align}
Together with $w_{\rm coh}=\epsc\nu_{\rm FC}$, this proves
Eq.~\eqref{eq:negativity-identity}.

Positivity of the initial density matrix requires
$|\rho_{ge}|\le\sqrt{\rho_{ee}\rho_{gg}}\le1/2$.  It follows from
Eq.~\eqref{eq:app-coherence-population-change} that
\begin{align}
 |\nu_{\rm FC}[\rho]|
 &\le
 2|\rho_{ge}|
 \sqrt{\mathcal T_{\rm ce}(1-\mathcal T_{\rm ce})}
 \nonumber\\
 &=C_{\ell_1}(\rho)
 \sqrt{\mathcal T_{\rm ce}(1-\mathcal T_{\rm ce})}
 \le\frac{C_{\ell_1}(\rho)}{2}.
 \label{eq:coherence-bound}
\end{align}
For the measurement-prepared state, this becomes
\begin{equation}
 \mathcal N_{\rm FC}
 \le |\sin\theta|
 \sqrt{\mathcal T_{\rm ce}(1-\mathcal T_{\rm ce})}
 \le\frac{|\sin\theta|}{2}.
 \label{eq:measurement-coherence-bound}
\end{equation}

Finally, for every integer $m\ge1$, the five weights give
\begin{align}
 \langle w^m\rangle_{\rm FC}
 &=(\epsc)^m\mathcal T_{\rm ce}
 \left[\rho_{gg}+(-1)^m\rho_{ee}\right]
 \nonumber\\
 &\quad+
 \left(\frac{\epsc}{2}\right)^m
 \left[1-(-1)^m\right]\nu_{\rm FC}[\rho].
 \label{eq:app-general-moment}
\end{align}
For even $m$, the interference pair cancels and
$\langle w^m\rangle_{\rm FC}=(\epsc)^m\mathcal T_{\rm ce}$.
For odd $m$,
\begin{equation}
 \langle w^m\rangle_{\rm FC}
 =(\epsc)^m\left[
 \mathcal T_{\rm ce}(\rho_{gg}-\rho_{ee})
 +2^{1-m}\nu_{\rm FC}[\rho]
 \right].
 \label{eq:app-general-odd-moment}
\end{equation}
Setting $\rho=\Pi_+$ and $m=2k$ or $2k+1$ proves
Eq.~\eqref{eq:moment-parity}.  The cases $m=1$ and $m=2$ give
Eq.~\eqref{eq:conditional-moments}.  Repeating the second-cumulant
calculation after setting $\nu_{\rm FC}=0$ while keeping the same populations
gives
\begin{equation}
 \kappa_{2,+}-\kappa_{2,+}^{\Delta}
 ={(\epsc)^2}\left(
 2\mathcal T_{\rm ce}\cos\theta\,\nu_{\rm FC}
 -\nu_{\rm FC}^2
 \right),
 \label{eq:app-coherence-cumulant-difference}
\end{equation}
which proves Eq.~\eqref{eq:coherence-cumulant-correction}.

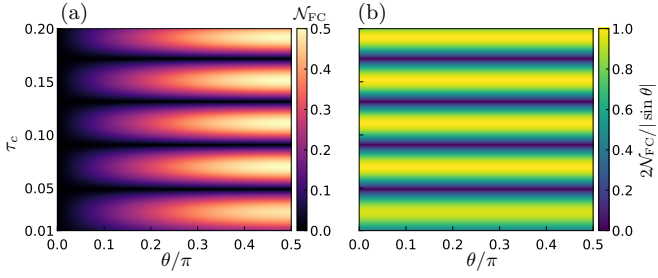
\begin{figure}[tbp]
 \centering 
  \hspace*{-0.04\columnwidth}
  \includegraphics[width=0.95\columnwidth]{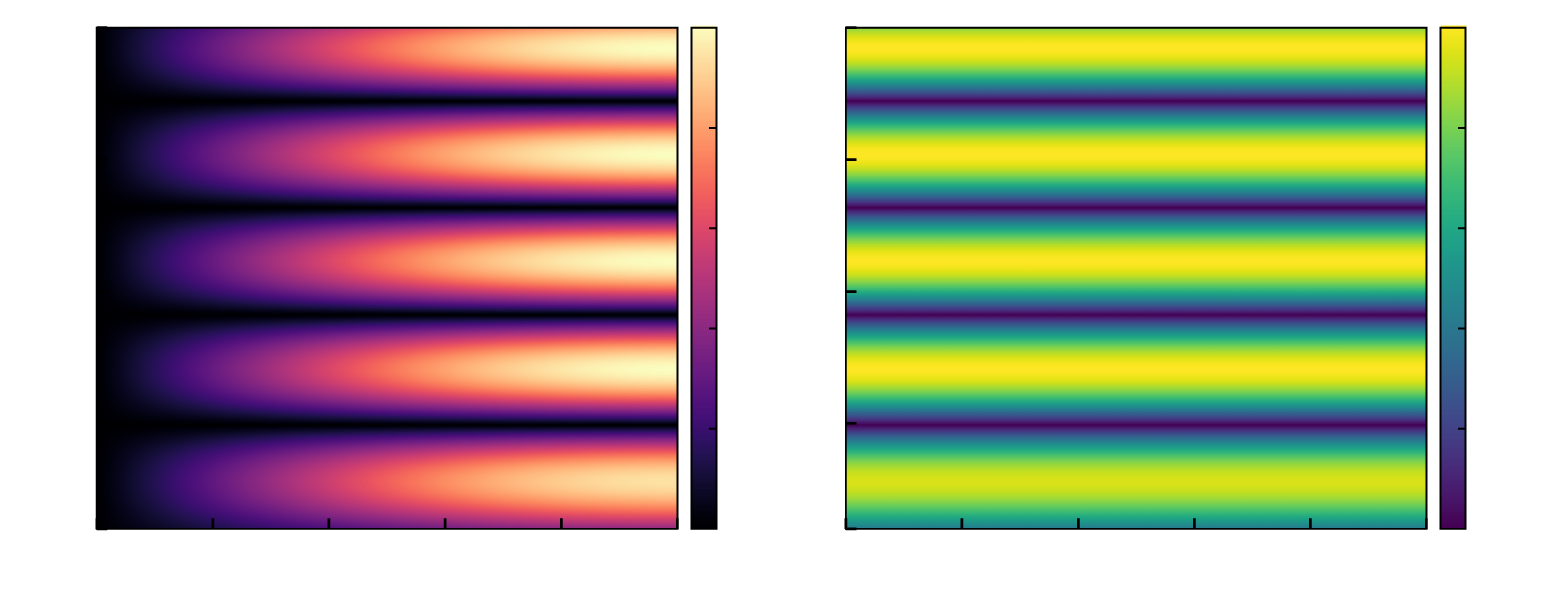}
  \put(-218,90){\scalebox{0.9}{(a)}}
   \put(-106,90){\scalebox{0.9}{(b)}}
  \put(-180,-5){\scalebox{0.8}{$\theta/\pi$}}
  \put(-65,-5){\scalebox{0.8}{$\theta/\pi$}}
 \put(-238,40){\rotatebox{90}{\scalebox{0.8}{$\tau_c$}}}
 \put(-130,90){{\scalebox{0.7}{$\mathcal N_{\rm FC}$}}}
 \put(0,33){\rotatebox{90}{\scalebox{0.7}{$2\mathcal N_{\rm FC}/|\sin\theta|$}}}
 \put(-220,3){\scalebox{0.7}{0}}
 \put(-205.5,3){\scalebox{0.7}{0.1}}
 \put(-188,3){\scalebox{0.7}{0.2}}
 \put(-171,3){\scalebox{0.7}{0.3}}
 \put(-153.5,3){\scalebox{0.7}{0.4}}
 \put(-136.5,3){\scalebox{0.7}{0.5}}
  \put(-233,9){\scalebox{0.7}{0.01}}
 \put(-233,24.5){\scalebox{0.7}{0.05}}
 \put(-230,44){\scalebox{0.7}{0.1}}
 \put(-233,65){\scalebox{0.7}{0.15}}
 \put(-230,83){\scalebox{0.7}{0.2}}
 \put(-122,9){\scalebox{0.7}{0}}
 \put(-123,22.5){\scalebox{0.7}{0.1}}
 \put(-123,39){\scalebox{0.7}{0.2}}
 \put(-123,54.2){\scalebox{0.7}{0.3}}
 \put(-123,68.8){\scalebox{0.7}{0.4}}
 \put(-123,83){\scalebox{0.7}{0.5}}
 \put(-105.5,3){\scalebox{0.7}{0}}
 \put(-90.8,3){\scalebox{0.7}{0.1}}
 \put(-73.3,3){\scalebox{0.7}{0.2}}
 \put(-55.8,3){\scalebox{0.7}{0.3}}
 \put(-38.3,3){\scalebox{0.7}{0.4}}
 \put(-21.4,3){\scalebox{0.7}{0.5}}
  \put(-8,9){\scalebox{0.7}{0}}
 \put(-9,22.5){\scalebox{0.7}{0.2}}
 \put(-9,39){\scalebox{0.7}{0.4}}
 \put(-9,54.2){\scalebox{0.7}{0.6}}
 \put(-9,68.8){\scalebox{0.7}{0.8}}
 \put(-9,83){\scalebox{0.7}{1}}

 \caption{Coherence--transition interference landscape.
 (a) Symmetric-FCS negativity versus measurement angle and compression
 duration at fixed $\tau_e=0.01$.
 (b) Ratio to the maximal coherence-only bound,
 $2\mathcal N_{\rm FC}/|\sin\theta|
 =2|\operatorname{Re}(U_{eg}U_{ee}^*)|\le1$.
 At $\theta=0$, the displayed ratio is defined by its continuous limit.
 The remaining parameters are $\omega_c=20\pi$, $\omega_h=30\pi$, and
 $\hbar=1$.}
 \label{fig:negativity-map}
\end{figure}

Figure~\ref{fig:negativity-map} shows the dependence of the resulting FCS
negativity on the measurement angle and compression duration.  The second
panel separates the available measurement coherence from the
drive-controlled interference factor.

\section{Outcome-resolved propagation over completed cycles}
\label{app:cycles}

The energy-basis construction of the outcome-resolved kernel in
Eq.~\eqref{eq:resolved-instrument} and the detailed derivations of the
finite-cycle generating function in Eq.~\eqref{eq:finite-N-generating}, the
two-mode representation in Eq.~\eqref{eq:two-mode-generating}, the
boundary--memory separation in Eq.~\eqref{eq:finite-N-asymptotic},
{and the exact finite-$N$ second cumulant in
Eq.~\eqref{eq:exact-binary-cumulant} are given below.}

\begin{figure}[h]
 \centering
 \includegraphics[width=0.98\columnwidth]{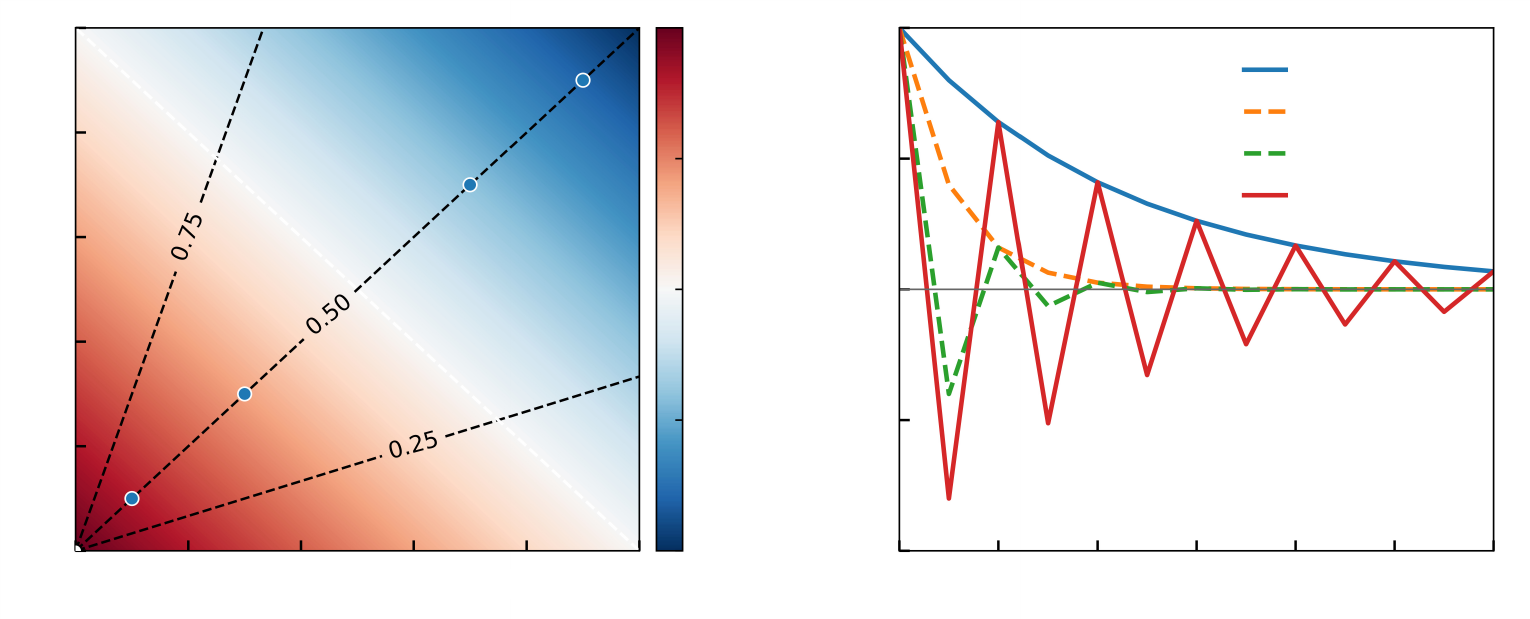}
 \put(-229,98){\scalebox{0.9}{(a)}}
 \put(-103,98){\scalebox{0.9}{(b)}}
 \put(-245,49){\rotatebox{90}{\scalebox{0.9}{$s$}}}
 \put(-186,-4){{\scalebox{0.9}{$r$}}}
 \put(-140,96){{\scalebox{0.8}{$\lambda_2$}}}
 \put(-58,-4){\scalebox{0.9}{$\ell$}}
 \put(-118,40){\rotatebox{90}{\scalebox{0.7}{$C_I(\ell)/C_I(0)$}}}
  \put(-230.5,3){\scalebox{0.7}{0}}
 \put(-215.5,3){\scalebox{0.7}{0.2}}
 \put(-198,3){\scalebox{0.7}{0.4}}
 \put(-181,3){\scalebox{0.7}{0.6}}
 \put(-163.5,3){\scalebox{0.7}{0.8}}
 \put(-143.3,3){\scalebox{0.7}{1}}
 \put(-236,9){\scalebox{0.7}{0}}
 \put(-240,24.3){\scalebox{0.7}{0.2}}
 \put(-240,41.5){\scalebox{0.7}{0.4}}
 \put(-240,58){\scalebox{0.7}{0.6}}
 \put(-240,74.2){\scalebox{0.7}{0.8}}
 \put(-236,89.9){\scalebox{0.7}{1}}
 \put(-131,10){\scalebox{0.7}{-1}}
 \put(-132,29.7){\scalebox{0.7}{-0.5}}
 \put(-131,49.5){\scalebox{0.7}{0}}
 \put(-132,70.2){\scalebox{0.7}{0.5}}
 \put(-131,89.9){\scalebox{0.7}{1}}
 \put(-100,3){\scalebox{0.7}{0}}
 \put(-85.5,3){\scalebox{0.7}{2}}
 \put(-70,3){\scalebox{0.7}{4}}
 \put(-54.8,3){\scalebox{0.7}{6}}
 \put(-39,3){\scalebox{0.7}{8}}
 \put(-25.3,3){\scalebox{0.7}{10}}
 \put(-10.3,3){\scalebox{0.7}{12}}
  \put(-107,10){\scalebox{0.7}{-1}}
 \put(-112,29.7){\scalebox{0.7}{-0.5}}
 \put(-106,49.5){\scalebox{0.7}{0}}
 \put(-110,70.2){\scalebox{0.7}{0.5}}
 \put(-107,89.9){\scalebox{0.7}{1}}
 \put(-37,86){{\scalebox{0.6}{$\lambda_2=0.8$}}}
 \put(-37,78.5){{\scalebox{0.6}{$\lambda_2=0.4$}}}
 \put(-37,72){{\scalebox{0.6}{$\lambda_2=-0.4$}}}
 \put(-37,65.6){{\scalebox{0.6}{$\lambda_2=-0.8$}}}
 \caption{Stationary composition and temporal memory of the binary outcome
 record.  (a) Subleading eigenvalue $\lambda_2=1-r-s$ over the
 transition-probability plane.  Black dashed contours show
 $p_+^*=0.25$, $0.50$, and $0.75$, and the white dashed line is the
 memoryless locus $\lambda_2=0$.  The four points on $r=s$ have the same
 stationary composition $p_+^*=1/2$ but different memory modes
 $\lambda_2=0.8$, $0.4$, $-0.4$, and $-0.8$.  The nonergodic origin is
 excluded.  (b) Normalized indicator covariance
 $C_I(\ell)/C_I(0)=\lambda_2^\ell$.  Positive modes are persistent, negative
 modes alternate, and equal $|\lambda_2|$ gives the same decay envelope.}
 \label{fig:markov-phase}
\end{figure}
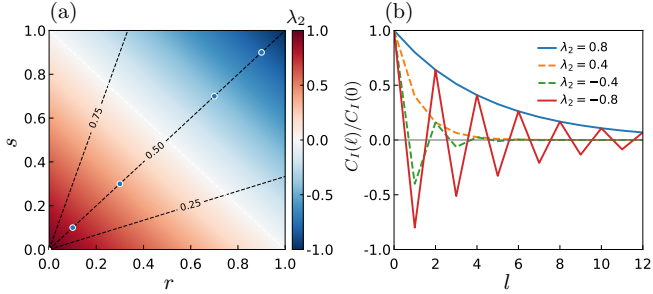

For the binary outcome map in Eq.~\eqref{eq:binary-outcome-map}, the
stationary composition in Eq.~\eqref{eq:stationary} depends on the ratio
$s/r$, whereas the subleading eigenvalue $\lambda_2$ in
Eq.~\eqref{eq:lambda2}, which governs the outcome memory, depends on their
sum.  Thus a fixed $p_+^*$ does not fix either the magnitude or the sign
of the temporal correlations.  For the driven-branch indicator,
Eq.~\eqref{eq:indicator-covariance} gives
$C_I(\ell)/C_I(0)=\lambda_2^\ell$.  Positive $\lambda_2$ produces persistent
records, negative $\lambda_2$ produces alternating records, {
and $\lambda_2=0$ gives independent measurement
outcomes}.  These distinctions are illustrated
in Fig.~\ref{fig:markov-phase}.

The work-resolved propagation is obtained by writing
$\Pi_b=|\psi_b\rangle\langle\psi_b|$ and expanding the driven column of
Eq.~\eqref{eq:resolved-instrument} in the endpoint energy bases.  With
$U_{\ell n}=\langle\ell|U_{\rm ce}|n\rangle$, the result is
\begin{multline}
 [\mathsf R(u)]_{b+}
 =\sum_{\ell,\ell',n,n'}
 \exp\!\left[\frac{\ii u}{2}
 \left(E_\ell^f+E_{\ell'}^f\right)\right]
 \\
 \times
 \exp\!\left[-\frac{\ii u}{2}
 \left(E_n^i+E_{n'}^i\right)\right]
 U_{\ell n}(\Pi_+)_{nn'}U_{\ell'n'}^*
 \\
 \times
 \langle\psi_b|\ell\rangle
 \langle\ell'|\psi_b\rangle.
 \label{eq:resolved-energy-basis}
\end{multline}
The thermal column is independent of $u$ because no driven work is assigned
to that branch.  Setting $u=0$ in
Eq.~\eqref{eq:resolved-energy-basis} gives
$[\mathsf R(0)]_{b+}=q_{b|+}$, and the corresponding relation for the
thermal column gives $\mathsf R(0)=R$.  Interchanging
$\ell\leftrightarrow\ell'$ and $n\leftrightarrow n'$ shows that
\begin{equation}
 \mathsf R(-u)=\mathsf R(u)^*.
 \label{eq:tilted-hermitian-symmetry}
\end{equation}

Summation over the next outcome uses $\sum_b\Pi_b=\id$ and yields
\begin{equation}
 \sum_b[\mathsf R(u)]_{b+}=\chi_+(u),
 \qquad
 \sum_b[\mathsf R(u)]_{b-}=1.
 \label{eq:resolved-column-sums}
\end{equation}
The sum over $b$ sets $\ell=\ell'$ and recovers the single-branch
characteristic function.  Before that sum is taken, the final outcome
projector couples the work field to the next outcome.  This is why the
factorization in Eq.~\eqref{eq:no-factorization} does not hold in general.

The resolved one-cycle weights are introduced through
\begin{equation}
 [\mathsf R(u)]_{ba}
 =\int dw\,\ee^{\ii uw}\mathcal Q_{ba}(w).
 \label{eq:resolved-work-weight}
\end{equation}
{Here, $\mathcal Q_{ba}(w)$ denotes the
outcome-resolved FCS work weight associated
with work $w$ and the transition from
outcome $a$ to outcome $b$ in one cycle.}
For the qubit, the integral denotes a finite sum of delta peaks.  The weights
in the driven column can be signed, but their outcome-summed normalization is
\begin{equation}
 \sum_b\int dw\,\mathcal Q_{ba}(w)=1.
 \label{eq:resolved-normalization}
\end{equation}
For an outcome path $a_0,a_1,\ldots,a_N$ and work increments
$w_0,\ldots,w_{N-1}$, the composed FCS weight is
\begin{equation}
 p_{a_0}^{(0)}
 \prod_{n=0}^{N-1}\mathcal Q_{a_{n+1}a_n}(w_n).
 \label{eq:path-weight}
\end{equation}
Multiplication by $\exp[\ii u\sum_n w_n]$, integration over the work
increments, and summation over all outcome paths give
\begin{align}
 \mathcal G_N(u)
 &=\sum_{a_0,\ldots,a_N}
 \prod_{n=0}^{N-1}[\mathsf R(u)]_{a_{n+1}a_n}
 p_{a_0}^{(0)}
 \nonumber\\
 &=\one^{\mathsf T}\mathsf R(u)^N\bm p^{(0)},
 \label{eq:app-finite-N-generating}
\end{align}
which proves Eq.~\eqref{eq:finite-N-generating}.  The matrix product is exact
for the selected FCS ordering, although the path weight in
Eq.~\eqref{eq:path-weight} is not required to be a positive classical
trajectory probability.

The binary kernel $M=\mathsf R(u)$ has trace $t=\Tr M$, determinant
$d=\det M$, and eigenvalues
\begin{equation}
 \Lambda_\pm(u)=\frac{t\pm\sqrt{t^2-4d}}{2}.
 \label{eq:tilted-eigenvalues}
\end{equation}
For an ergodic record, the two eigenvalues are distinct in a neighborhood of
$u=0$ because $\Lambda_+(0)=1$ and
$\Lambda_-(0)=\lambda_2$ with $|\lambda_2|<1$.  The
Cayley--Hamilton identity gives, for $N\ge1$,
\begin{multline}
 M^N=\frac{1}{\Lambda_+-\Lambda_-}
 \left[\Lambda_+^N(M-\Lambda_-\id)\right.\\
 \left.{}-\Lambda_-^N(M-\Lambda_+\id)\right].
 \label{eq:finite-matrix-power}
\end{multline}
Contracting Eq.~\eqref{eq:finite-matrix-power} with
$\one^{\mathsf T}$ and $\bm p^{(0)}$ proves
Eq.~\eqref{eq:two-mode-generating}, with
\begin{align}
 A_+(u)&=\frac{g_1(u)-\Lambda_-(u)}
 {\Lambda_+(u)-\Lambda_-(u)},
 \nonumber\\
 A_-(u)&=\frac{\Lambda_+(u)-g_1(u)}
 {\Lambda_+(u)-\Lambda_-(u)},
 \nonumber\\
 g_1(u)&=\one^{\mathsf T}\mathsf R(u)\bm p^{(0)}.
 \label{eq:mode-amplitudes}
\end{align}

Near $u=0$, set
$\psi(u)=\ln\Lambda_+(u)$,
$B(u)=A_-(u)/A_+(u)$, and
$r(u)=\Lambda_-(u)/\Lambda_+(u)$.  Taking the logarithm of the two-mode
result gives
\begin{equation}
 \ln\mathcal G_N(u)
 =N\psi(u)+\ln A_+(u)
 +\ln\!\left[1+B(u)r(u)^N\right].
 \label{eq:bulk-boundary-memory}
\end{equation}
Differentiation yields
\begin{align}
 \kappa_m^{(N)}
 &=N\kappa_{m,\infty}+\mathcal B_m+\delta_{m,N},
 \nonumber\\
 \mathcal B_m
 &=(-\ii)^m\left.\partial_u^m\ln A_+(u)\right|_{u=0},
 \nonumber\\
 \delta_{m,N}
 &=(-\ii)^m\left.\partial_u^m
 \ln[1+B(u)r(u)^N]\right|_{u=0}.
 \label{eq:finite-cumulant-separation}
\end{align}
At the origin, $A_+(0)=1$, $A_-(0)=0$, and $r(0)=\lambda_2$.
Analyticity therefore implies, for fixed $m$ and any
$|\lambda_2|<\varrho<1$,
\begin{equation}
 \delta_{m,N}
 =\mathcal O\!\left(N^{m-1}\varrho^N\right),
 \label{eq:memory-remainder}
\end{equation}
which proves Eq.~\eqref{eq:finite-N-asymptotic}.  The second term in
Eq.~\eqref{eq:finite-cumulant-separation} is an $N$-independent boundary
contribution, while the last term carries the decaying outcome-memory mode.

For stationary initialization, the first cumulant is exactly extensive.
Indeed,
\begin{align}
 \left.\partial_u\mathcal G_N(u)\right|_{u=0}
 &=\sum_{j=0}^{N-1}\one^{\mathsf T}
 \mathsf R'(0)R^j\bm p^*
 \nonumber\\
 &=N\one^{\mathsf T}\mathsf R'(0)\bm p^*,
 \label{eq:stationary-mean-proof}
\end{align}
where $R\bm p^*=\bm p^*$ and
$\one^{\mathsf T}R=\one^{\mathsf T}$ have been used.  Higher cumulants
generically retain both terms in
Eq.~\eqref{eq:finite-cumulant-separation}.

A compact calculation of the long-cycle second cumulant follows from the
stationary eigenvalue of a real moment kernel.  Define
\begin{equation}
 \mathsf M(k)=\mathsf R(-\ii k),
 \qquad
 \mathsf M_n=\left.\partial_k^n\mathsf M(k)\right|_{k=0},
 \label{eq:moment-tilted-matrices}
\end{equation}
and introduce the projector and group inverse
\begin{equation}
 \mathcal P_*=\bm p^*\one^{\mathsf T},
 \qquad
 \mathsf G=(\id-R+\mathcal P_*)^{-1}-\mathcal P_*.
 \label{eq:app-group-inverse}
\end{equation}
Let $\lambda(k)$ be the eigenvalue of $\mathsf M(k)$ satisfying
$\lambda(0)=1$.  Standard eigenvalue perturbation gives
\cite{Meyer1975}
\begin{align}
 \lambda'(0)
 &=\one^{\mathsf T}\mathsf M_1\bm p^*,
 \nonumber\\
 \lambda''(0)
 &=\one^{\mathsf T}\mathsf M_2\bm p^*
 +2\one^{\mathsf T}\mathsf M_1\mathsf G\mathsf M_1\bm p^*.
 \label{eq:app-eigenvalue-derivatives}
\end{align}
Since the scaled cumulant generator is $\ln\lambda(k)$, it follows that
\begin{align}
 \bar w
 &=\one^{\mathsf T}\mathsf M_1\bm p^*,
 \nonumber\\
 \kappa_{2,\infty}
 &=\one^{\mathsf T}\mathsf M_2\bm p^*-\bar w^2
 +2\one^{\mathsf T}\mathsf M_1\mathsf G\mathsf M_1\bm p^*
 \nonumber\\
 &=\kappa_{2,1}
 +2\one^{\mathsf T}\mathsf M_1\mathsf G\mathsf M_1\bm p^*.
 \label{eq:app-group-second}
\end{align}
For this engine, the first term is precisely
the stationary one-cycle baseline in Eq.~\eqref{eq:one-cycle-cumulants}.
{
The binary outcome process admits the spectral
decomposition
$R=\mathcal P_*+\lambda_2(\id-\mathcal P_*)$, yielding
$\mathsf G=(\id-\mathcal P_*)/(1-\lambda_2)$.
Defining
$A=\mathbf{1}^{\mathsf T}\mathsf M_1(\id-\mathcal P_*)\mathsf M_1\bm p^*$,
the connected FCS work moment at cycle separation
$\ell\geq1$ is
}
{
\begin{align}
C_w(\ell)
&=\mathbf{1}^{\mathsf T}\mathsf M_1
(R^{\ell-1}-\mathcal P_*)\mathsf M_1\bm p^*
\nonumber\\
&=A\lambda_2^{\ell-1}.
\label{eq:work-correlation-binary}
\end{align}
}
{
Thus, $C_w(1)=A$ can remain nonzero at
$\lambda_2=0$, whereas $C_w(\ell)=0$
for $\ell\geq2$.
These are connected moments of the FCS construction.
For stationary initialization, the exact finite-cycle
second cumulant is
}
{
\begin{equation}
\kappa_2^{(N)}
=N\kappa_{2,1}
+2\sum_{\ell=1}^{N-1}(N-\ell)C_w(\ell).
\label{eq:work-cumulant-sum}
\end{equation}
}
{
Substituting Eq.~\eqref{eq:work-correlation-binary}
into Eq.~\eqref{eq:work-cumulant-sum} and evaluating
the geometric sum yields
Eq.~\eqref{eq:exact-binary-cumulant}
of the main text.
}

\section{Fixed-time work-statistics derivation}
\label{app:physical-time}

The root relation in Eq.~\eqref{eq:physical-time-root} and the work-rate
formulas in Eqs.~\eqref{eq:mean-work-rate}
and~\eqref{eq:physical-time-second-cumulant} are derived here for the
deterministic branch durations in Eq.~\eqref{eq:branch-durations}.

Consider an outcome path $a_0,a_1,\ldots,a_N$.  If $w_n$ is the FCS work
increment associated with the completed transition $a_n\to a_{n+1}$, the
work and elapsed time after $N$ completed cycles are
\begin{equation}
 W_N=\sum_{n=0}^{N-1}w_n,
 \qquad
 T_N=\sum_{n=0}^{N-1}\tau_{a_n}.
 \label{eq:app-fixed-path}
\end{equation}
The transform factor for a completed branch $a$ is therefore
$\exp(\ii u w-\zeta\tau_a)$.  Combining this factor with the
outcome-resolved work kernel in Eq.~\eqref{eq:resolved-instrument} gives
\begin{equation}
 [\mathsf R(u,\zeta)]_{ba}
 =\ee^{-\zeta\tau_a}[\mathsf R(u)]_{ba},
 \label{eq:app-fixed-joint-kernel}
\end{equation}
which is the column-weighted kernel
$\mathsf R(u,\zeta)=\mathsf R(u)\mathsf D(\zeta)$ introduced in
Eq.~\eqref{eq:joint-kernel}.  Matrix composition then yields
\begin{equation}
 \mathcal G_N(u,\zeta)
 =\one^{\mathsf T}\mathsf R(u,\zeta)^N\bm p^{(0)}
 =\left\langle\ee^{\ii uW_N-\zeta T_N}\right\rangle_{\rm FC}.
 \label{eq:app-fixed-completed}
\end{equation}

At laboratory time $t$, let
$N(t)=\max\{N:T_N\le t\}$ and set $W(t)=W_{N(t)}$, so that only completed
branches contribute to the counted work.  If the last completed transition
leaves the process in outcome $a$, the age $x$ of the unfinished branch lies
in $0\le x<\tau_a$.  Its Laplace transform is
\begin{equation}
 \mathcal S_a(\zeta)
 =\int_0^{\tau_a}\!dx\,\ee^{-\zeta x}
 =\frac{1-\ee^{-\zeta\tau_a}}{\zeta}.
 \label{eq:app-fixed-survival}
\end{equation}
Writing these factors as the vector $\bm{\mathcal S}(\zeta)$ and summing over
the number of completed cycles gives
\begin{align}
 \widetilde{\mathcal Z}(u,\zeta)
 &\equiv
 \int_0^\infty\!dt\,\ee^{-\zeta t}
 \left\langle\ee^{\ii uW(t)}\right\rangle_{\rm FC}
 \nonumber\\
 &=\sum_{N=0}^{\infty}
 \bm{\mathcal S}(\zeta)^{\mathsf T}
 \mathsf R(u,\zeta)^N\bm p^{(0)}
 \nonumber\\
 &=\bm{\mathcal S}(\zeta)^{\mathsf T}
 [\id-\mathsf R(u,\zeta)]^{-1}\bm p^{(0)}.
 \label{eq:app-fixed-resolvent}
\end{align}
This expression is normalized correctly.  Indeed,
$\bm{\mathcal S}(\zeta)^{\mathsf T}
=\one^{\mathsf T}[\id-\mathsf D(\zeta)]/\zeta$ and
\(
 \one^{\mathsf T}[\id-R\mathsf D(\zeta)]
 =\one^{\mathsf T}[\id-\mathsf D(\zeta)]
\), so Eq.~\eqref{eq:app-fixed-resolvent} gives
$\widetilde{\mathcal Z}(0,\zeta)=1/\zeta$.

The extensive fixed-time statistics are determined by the pole of the
resolvent that evolves continuously from the stationary pole at the origin.
Let $\Lambda_0(u,\zeta)$ denote the eigenvalue of
$\mathsf R(u,\zeta)$ satisfying $\Lambda_0(0,0)=1$.  The outcome chain is
assumed ergodic, so this eigenvalue is simple at the origin.  Its time
derivative is
\(
 \partial_\zeta\Lambda_0(0,0)=-\bar\tau<0
\), where
$\bar\tau=p_+^*\tau_++p_-^*\tau_-$.  The implicit-function theorem therefore
gives a unique analytic pole $\zeta=\phi(u)$ near the origin, determined
by
\begin{equation}
 \Lambda_0\!\left(u,\phi(u)\right)=1,
 \qquad
 \phi(0)=0.
 \label{eq:app-fixed-root}
\end{equation}
The remaining factors in Eq.~\eqref{eq:app-fixed-resolvent} affect only the
pole residue.  Inverse Laplace transformation consequently gives
\begin{equation}
 \ln\left\langle\ee^{\ii uW(t)}\right\rangle_{\rm FC}
 =t\phi(u)+\mathcal O(1).
 \label{eq:app-fixed-asymptotic}
\end{equation}
Equation~\eqref{eq:app-fixed-root} is precisely the main-text root relation
in Eq.~\eqref{eq:physical-time-root}.

It remains to differentiate this root.  Define
$\Psi(u,\zeta)=\ln\Lambda_0(u,\zeta)$.  Equation~\eqref{eq:app-fixed-root}
is equivalent to
$\Psi(u,\phi(u))=0$.  Its first derivative at $u=0$ is
\begin{equation}
 \phi'(0)
 =-\frac{\Psi_u}{\Psi_\zeta}
 =\ii\frac{\bar w}{\bar\tau},
 \label{eq:app-fixed-first-derivative}
\end{equation}
where the derivatives are evaluated at $(u,\zeta)=(0,0)$.  Hence
$j_w\equiv-\ii\phi'(0)=\bar w/\bar\tau$, which proves
Eq.~\eqref{eq:mean-work-rate}.

A second differentiation gives
\begin{equation}
 \phi''(0)
 =-\frac{
 \Psi_{uu}+2\Psi_{u\zeta}\phi'(0)
 +\Psi_{\zeta\zeta}[\phi'(0)]^2}
 {\Psi_\zeta}.
 \label{eq:app-fixed-second-derivative}
\end{equation}
The Fourier--Laplace convention
$\exp(\ii uW-\zeta T)$ and the definitions in
Eq.~\eqref{eq:joint-cumulants} imply
\begin{align}
 \Psi_u&=\ii\bar w,
 &\Psi_\zeta&=-\bar\tau,\nonumber\\
 \Psi_{uu}&=-C_{ww},
 &\Psi_{u\zeta}&=-\ii C_{w\tau},
 &\Psi_{\zeta\zeta}&=C_{\tau\tau}.
 \label{eq:app-fixed-dictionary}
\end{align}
Substitution of Eq.~\eqref{eq:app-fixed-dictionary} and
$\phi'(0)=\ii j_w$ into
Eq.~\eqref{eq:app-fixed-second-derivative} yields
\begin{equation}
 -\phi''(0)
 =\frac{C_{ww}-2j_wC_{w\tau}+j_w^2C_{\tau\tau}}
 {\bar\tau}.
 \label{eq:app-fixed-second-rate}
\end{equation}
This proves Eq.~\eqref{eq:physical-time-second-cumulant}.  Thus the
completed-cycle fluctuation, the work--duration cross correlation, and the
duration fluctuation all enter when work is sampled at a fixed laboratory
time.

\begin{figure}[tbp]
\vspace*{0.3cm}
 \centering
 \includegraphics[width=\columnwidth]{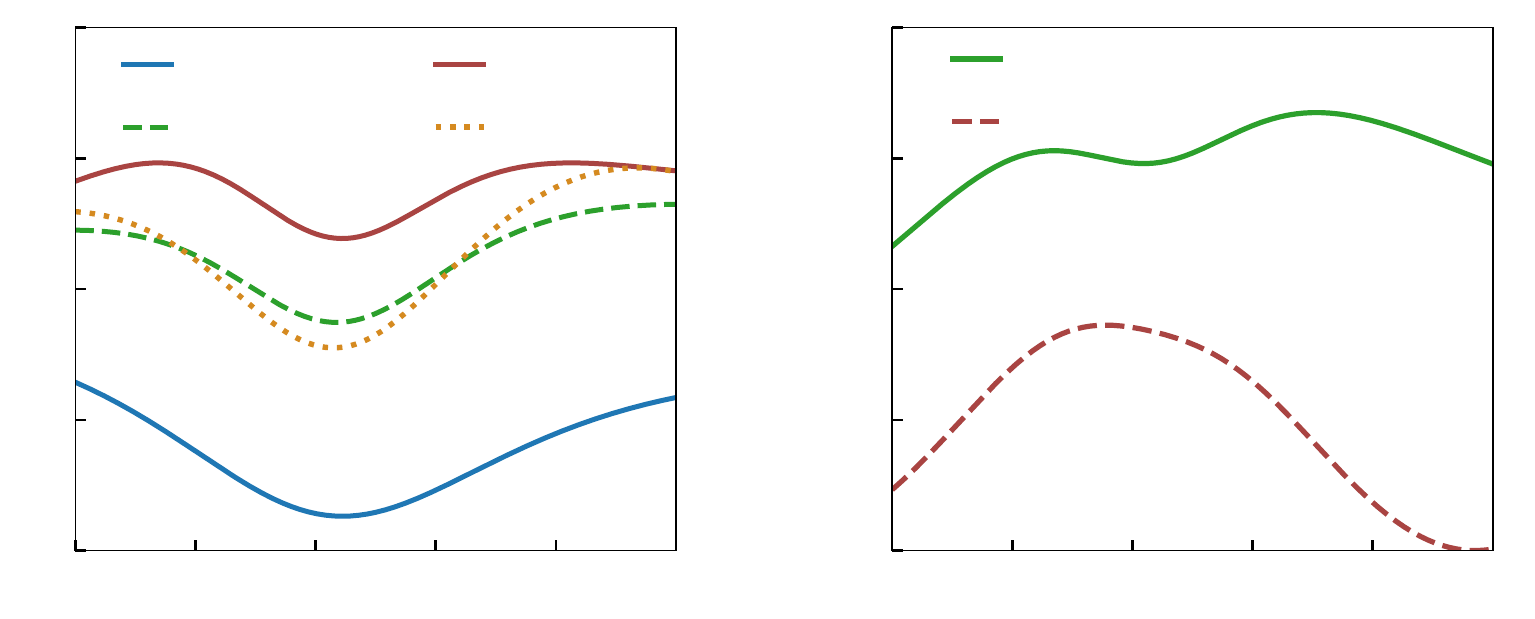}
 \put(-234,100){\scalebox{0.9}{(a)}}
 \put(-103,100){\scalebox{0.9}{(b)}}
 \put(-216,88){\scalebox{0.5}{$Q_{S|X}/(\kb T)$}}
 \put(-216,78){\scalebox{0.5}{$Q_{S}/(\kb T)$}}
 \put(-165,88){\scalebox{0.5}{$Q_X^{(1)}/(\kb T)$}}
 \put(-165,78){\scalebox{0.5}{$Q_X^{(\infty)}/(\kb T)$}}
  \put(-83,89){\scalebox{0.6}{$\chi_{X:S}$}}
  \put(-83,78){\scalebox{0.6}{$I(X_n:X_{n-1})$}}
  \put(-189,-5){\scalebox{0.8}{$\theta/\pi$}}
  \put(-59,-5){\scalebox{0.8}{$\theta/\pi$}}
  \put(-124,10){\rotatebox{90}{\scalebox{0.7}{\text{side-information saving (nats)}}}}
  \put(-234,3){\scalebox{0.7}{0}}
 \put(-218.5,3){\scalebox{0.7}{0.1}}
 \put(-198.8,3){\scalebox{0.7}{0.2}}
 \put(-181,3){\scalebox{0.7}{0.3}}
 \put(-163,3){\scalebox{0.7}{0.4}}
 \put(-143.3,3){\scalebox{0.7}{0.5}}
 \put(-244,10){\scalebox{0.7}{0.1}}
 \put(-244,29.7){\scalebox{0.7}{0.3}}
 \put(-244,50){\scalebox{0.7}{0.5}}
 \put(-244,71.2){\scalebox{0.7}{0.7}}
 \put(-244,91.2){\scalebox{0.7}{0.9}}
 \put(-104,3){\scalebox{0.7}{0}}
 \put(-87,3){\scalebox{0.7}{0.1}}
 \put(-67.7,3){\scalebox{0.7}{0.2}}
 \put(-48.5,3){\scalebox{0.7}{0.3}}
 \put(-29.5,3){\scalebox{0.7}{0.4}}
 \put(-9.3,3){\scalebox{0.7}{0.5}}
  \put(-110,10){\scalebox{0.7}{0}}
 \put(-113,29.7){\scalebox{0.7}{0.1}}
 \put(-113,50){\scalebox{0.7}{0.2}}
 \put(-113,71.2){\scalebox{0.7}{0.3}}
 \put(-113,91.2){\scalebox{0.7}{0.4}}
 \caption{Distinct controller reset ledgers.
 (a) Conditional-state charge $Q_{S|X}$, state-only charge $Q_S$,
 single-symbol outcome charge $Q_X^{(1)}$, and long-record outcome charge
 $Q_X^{(\infty)}$.
 (b) Holevo saving $\chi_{X:S}$ and predictive saving
 $I(X_n:X_{n-1})$.  The two savings concern different physical memories.
 The parameters are those specified in Sec.~II, {with
$\tau_d=2$.}}
 \label{fig:reset}
\end{figure}
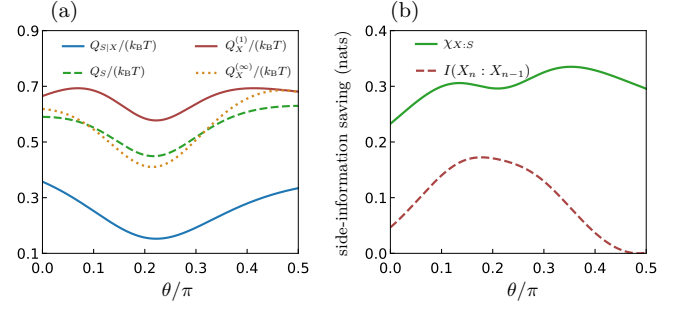

\section{Record-reset identities and physical ledgers}
\label{app:reset}

The entropy chain rule is used below to derive
Eqs.~\eqref{eq:block-entropy},
\eqref{eq:block-cost}, and
\eqref{eq:excess-entropy} of the main text.  The classical--quantum archive
construction is then used to derive Eq.~\eqref{eq:joint-cost}.  The remaining
relations distinguish alternative physical reset ledgers; they are not costs
to be added unless the corresponding memories are all present and erased.

\begin{table*}[t]
 \caption{Minimum reversible reset charges per recorded cycle for degenerate
 logical Hamiltonians.  The state register $S$ is included only when it is
 physically retained in addition to the outcome register $X$.}
 \label{tab:reset-ledgers}
 \small
 \begin{ruledtabular}
 \begin{tabular}{@{}p{0.26\textwidth}p{0.68\textwidth}@{}}
 Physical reset ledger & Minimum charge per cycle\\
 \colrule
 Single outcome $X$
  & $Q_X^{(1)}=\kb T H_2(p_+^*)$\\
Length-$N$ record $X^N$
  & $Q_X^{(N)}/N=\kb T h_\mu
  +\kb T[H_2(p_+^*)-h_\mu]/N$\\
 Conditional state $S|X$
  & $Q_{S|X}=\kb T\sum_a p_a^*S(\rho_a)$\\
 Unconditioned state $S$
  & $Q_S=\kb T S(\bar\rho)$\\
 Joint archive $X^NS^N$, $N\to\infty$
  & $Q_{XS}^{(\infty)}=Q_X^{(\infty)}+Q_{S|X}$\\
 \end{tabular}
 \end{ruledtabular}
\end{table*}

For a stationary first-order Markov record
$X^N=(X_1,\ldots,X_N)$, the path probability is
\begin{equation}
 p(a_1,\ldots,a_N)
 =p_{a_1}^*\prod_{n=1}^{N-1}q_{a_{n+1}|a_n}.
 \label{eq:app-record-probability}
\end{equation}
The entropy chain rule gives
\begin{align}
 H(X^N)
 &=H(X_1)+\sum_{n=2}^{N}H(X_n|X_1,\ldots,X_{n-1})
 \nonumber\\
 &=H(X_1)+\sum_{n=2}^{N}H(X_n|X_{n-1}),
 \label{eq:app-record-chain-rule}
\end{align}
where the Markov property has been used in the second line.  Stationarity
implies
\begin{equation}
 H(X_1)=H_2(p_+^*),
 \qquad
 H(X_n|X_{n-1})=h_\mu,
 \label{eq:app-stationary-record-entropies}
\end{equation}
with $h_\mu$ defined in Eq.~\eqref{eq:entropy-rate}.  Substitution into
Eq.~\eqref{eq:app-record-chain-rule} yields
\begin{equation}
 H(X^N)=H_2(p_+^*)+(N-1)h_\mu,
 \label{eq:app-record-block-entropy}
\end{equation}
which proves Eq.~\eqref{eq:block-entropy}.  For degenerate logical states,
reversible isothermal erasure at temperature $T$ requires a minimum charge
$\kb T H(X^N)$.  Division by $N$ in
Eq.~\eqref{eq:app-record-block-entropy} then proves
Eq.~\eqref{eq:block-cost}.

The mutual information between two consecutive stationary outcomes is
\begin{align}
 I(X_n:X_{n-1})
 &=H(X_n)-H(X_n|X_{n-1})\nonumber\\
 &=H_2(p_+^*)-h_\mu.
 \label{eq:app-predictive-information}
\end{align}
Equation~\eqref{eq:excess-entropy} follows immediately.  Thus the exact
$1/N$ term in Eq.~\eqref{eq:block-cost} is the predictive information stored
at the initial boundary of the block.

The physical memory carrying the outcome record must be distinguished from
any separately retained state register.  Let $X$ denote the classical
outcome register and let $S$ denote an optional state-bearing register.  If
both are retained after one stationary cycle, their classical--quantum state
is
\begin{equation}
 \Omega_{XS}
 =\sum_{a=\pm}p_a^*|a\rangle\langle a|_X\otimes\rho_a.
 \label{eq:cq-state}
\end{equation}
If $X$ remains available as side information while $S$ is reset, the entropy
removed from $S$ is the conditional entropy of this state, giving
\begin{equation}
 Q_{S|X}=\kb T\sum_a p_a^*S(\rho_a),
 \qquad
 S(\rho)=-\Tr(\rho\ln\rho).
 \label{eq:conditional-state-cost}
\end{equation}
Since $\rho_+$ is a unitary image of the rank-one state $\Pi_+$,
$Q_{S|X}=\kb T p_-^*S(\rho_-)$ for the present protocol
\cite{companion,Liu25}.  If $S$ is instead reset without access to $X$, its
average state $\bar\rho=\sum_a p_a^*\rho_a$ must be erased, and the charge is
\begin{equation}
 Q_S=\kb T S(\bar\rho).
 \label{eq:state-only-cost}
\end{equation}
Their difference is the Holevo information \cite{Holevo1973},
\begin{align}
 Q_S-Q_{S|X}&=\kb T\chi_{X:S},\nonumber\\
 \chi_{X:S}
 &=S(\bar\rho)-\sum_a p_a^*S(\rho_a)\ge0.
 \label{eq:app-holevo-saving}
\end{align}

For the classical outcome register alone, symbol-by-symbol erasure gives
\begin{equation}
 Q_X^{(1)}=\kb T H_2(p_+^*),
 \label{eq:one-symbol-cost}
\end{equation}
whereas Eqs.~\eqref{eq:app-record-block-entropy} and
\eqref{eq:block-cost} give the asymptotic per-cycle charge
\begin{equation}
 Q_X^{(\infty)}=\kb T h_\mu\le Q_X^{(1)}.
 \label{eq:long-record-cost}
\end{equation}
The reduction
$Q_X^{(1)}-Q_X^{(\infty)}=\kb T I(X_n:X_{n-1})$
is associated with correlations in the outcome record.  It is distinct from
the Holevo reduction in Eq.~\eqref{eq:app-holevo-saving}, which uses the
outcome as side information for a state register.

 The exact $1/N$ dependence follows directly
from Eq.~\eqref{eq:block-cost}; no additional asymptotic approximation is
used.
Figure~\ref{fig:reset} compares the alternative ledgers and the two distinct
side-information savings.  The figure is a comparison of controller
architectures, not a sum of reset costs.

For completeness, suppose that a length-$N$ outcome record and $N$ separate
state registers are jointly archived.  Conditioned on the outcome sequence,
the state registers are assumed to have the product state
$\rho_{a_1}\otimes\cdots\otimes\rho_{a_N}$.  The archive state is then
\begin{align}
 \Omega_{X^NS^N}
 &=\sum_{a_1,\ldots,a_N}p(a_1,\ldots,a_N)
 |a_1\cdots a_N\rangle\langle a_1\cdots a_N|_{X^N}
 \nonumber\\
 &\qquad\otimes
 \rho_{a_1}\otimes\cdots\otimes\rho_{a_N}.
 \label{eq:app-joint-archive}
\end{align}
The entropy formula for a block-diagonal classical--quantum state and the
additivity of entropy for product states give
\begin{align}
 S(\Omega_{X^NS^N})
 &=H(X^N)
 +\sum_{a_1,\ldots,a_N}p(a_1,\ldots,a_N)
 \sum_{n=1}^{N}S(\rho_{a_n})\nonumber\\
 &=H(X^N)+N\sum_a p_a^*S(\rho_a),
 \label{eq:app-joint-archive-entropy}
\end{align}
where stationarity has been used in the second line.  Multiplication by
$\kb T/N$, followed by the limit $N\to\infty$, yields
\begin{equation}
 \lim_{N\to\infty}
 \frac{\kb T}{N}S(\Omega_{X^NS^N})
 =\kb T\left[h_\mu+\sum_a p_a^*S(\rho_a)\right],
 \label{eq:app-joint-archive-cost}
\end{equation}
which proves Eq.~\eqref{eq:joint-cost}.

The resulting reset ledgers are summarized in
Table~\ref{tab:reset-ledgers}.  Each row refers to a different physical
boundary of the controller.

\end{document}